\documentclass[superscriptaddress,amsmath,amssymb,prd,twocolumn,nofootinbib]{revtex4-2}
\usepackage{natbib}
\usepackage{graphics}
\usepackage{multirow}
\usepackage{amsbsy}
\usepackage{footmisc}
\usepackage{hyperref}
\usepackage{amsmath}
\usepackage{aas_macros}

\usepackage{array}
\usepackage[mathscr]{eucal}
\usepackage{morefloats}
\usepackage{amssymb}
\usepackage{txfonts}
\usepackage{placeins}
\usepackage{natbib}

\usepackage{dcolumn}
\usepackage{makecell}
\usepackage{booktabs}

\AtBeginDocument{
\heavyrulewidth=.08em
\lightrulewidth=.05em
\cmidrulewidth=.03em
\belowrulesep=.65ex
\belowbottomsep=0pt
\aboverulesep=.4ex
\abovetopsep=0pt
\cmidrulesep=\doublerulesep
\cmidrulekern=.5em
\defaultaddspace=.5em
}

\DeclareMathAlphabet{\mathbfsf}{\encodingdefault}{\sfdefault}{bx}{n}

\usepackage{verbatim}
\usepackage{graphicx}
\usepackage{epsf}
\usepackage{subfigure}
\usepackage{color}
\usepackage{threeparttable}
\usepackage{comment}
\usepackage{epsfig}
\usepackage{xspace}
\usepackage{enumitem}
\usepackage{hyperref}
\usepackage{ulem}
\usepackage{courier}
\usepackage{appendix}
\usepackage{comment}
\usepackage[usenames,dvipsnames,svgnames]{xcolor}
\usepackage{latexsym}
\usepackage{url}

\hypersetup{
  colorlinks = true,
  urlcolor = blue,
  linkcolor=blue,
  citecolor=blue
}

\usepackage[applemac]{inputenc}

\def\planck{{Planck}}

\def\spt{{\sc SPT}}
\def\SPT{{\spt}}
\def\sptyr{{\sc SPT-3G 2018}}
\def\actdr{{\sc ACT DR4}}
\def\SPTyr{\sptyr}
\def\b{{\it b}}

\def\actlite{{\tt ACTPollite}}

\def\actdrlite{{\tt actpollite dr4}}

\def\act{{\sc ACT}}
\def\ACT{{\act}}

\newcommand{\mksym}[1]{\ifmmode {\rm #1}\else #1\fi}

\newcommand{\TT}{\mksym{TT}}
\newcommand{\TE}{\mksym{TE}}
\newcommand{\EE}{\mksym{EE}}
\newcommand{\TEEE}{\TE,\EE}

\newcommand{\TTTEEE}{{TT,TE,EE}}

\providecommand{\omb}{\omega_{\mathrm{b}}}
\providecommand{\omc}{\omega_{\mathrm{c}}}

\newcommand{\ns}{n_{\rm s}}

\newcommand{\As}{A_{\rm s}}
\newcommand{\clamp}{\As \mathrm{e}^{-2\tau}}

\newcommand{\lcdm}{\texorpdfstring{{$\rm{\Lambda CDM}$}}{ΛCDM}}

\newcommand{\LCDM}{\lcdm}
\newcommand{\lcdmb}{\texorpdfstring{{${\rm\Lambda CDM+}{\it b}$ }}{ΛCDM}}

\newcommand{\lnAs}{\ln(10^{10} A_{\rm s})}

\newcommand{\nnu}{N_{\rm eff}}
\newcommand{\yp}{y_{p}}
\newcommand{\ypTE}{y^{TE}_{p}}
\newcommand{\Astau}{\As e^{-2\tau}}

\begin{document}
\title{Consistency of Planck, ACT and SPT constraints on magnetically assisted recombination and forecasts for future experiments}

\author{\href{https://orcid.org/0000-0002-1399-5057}{Silvia Galli}}
\affiliation{Institut d'Astrophysique de Paris, UMR 7095, CNRS \& Sorbonne Universit\'{e}, 98 bis boulevard Arago, 75014 Paris, France}
\email[]{gallis@iap.fr}

\author{\href{https://orcid.org/0000-0001-5108-0854}{Levon Pogosian}} 
\affiliation{Department of Physics, Simon Fraser University, Burnaby, BC, V5A 1S6, Canada}
\email[]{levon@sfu.ca}

\author{Karsten Jedamzik} 
\affiliation{Laboratoire de Univers et Particules de Montpellier, UMR5299-CNRS, Universite de Montpellier, 34095 Montpellier, France}
\email[]{karsten.jedamzik@umontpellier.fr}

\author{\href{https://orcid.org/0000-0001-6899-1873}{Lennart Balkenhol}}
\affiliation{School of Physics, University of Melbourne, Parkville, VIC 3010, Australia}
\email[]{lbalkenhol@student.unimelb.edu.au}

\date{\today}
\begin{abstract}
Primordial magnetic fields can change the recombination history of the universe by inducing clumping in the baryon density at small scales. They were recently proposed as a candidate model to relieve the Hubble tension. We investigate the consistency of the constraints on a clumping factor parameter $b$ in a simplistic model, using the latest CMB data from Planck, ACT DR4 and SPT-3G 2018. 
For the combined CMB data alone, we find no evidence for clumping being different from zero, though when adding a prior on $H_0$ based on the latest distance-ladder analysis of the SH0ES team, we report a weak detection of $b$. Our analysis of simulated datasets shows that ACT DR4 has more constraining power with respect to SPT-3G 2018 due to the degeneracy breaking power of the TT band powers (not included in SPT). Simulations also suggest that the TE,EE power spectra of the two datasets should have the same constraining power. However, the ACT DR4 TE,EE constraint is tighter than expectations, while the SPT-3G 2018 one is looser. While this is compatible with statistical fluctuations, we explore systematic effects which may account for such deviations.
Overall, the ACT results are only marginally consistent with Planck or SPT-3G, at the $2-3\sigma$ level within $\Lambda$CDM+$b$ and $\Lambda$CDM, while Planck and SPT-3G are in good agreement. Combining the CMB data together with BAO and SNIa provides an upper limit of $b<0.4$ at 95\% c.l. ($b<0.5$ without ACT). Adding a SH0ES-based prior on the Hubble constant gives $b = 0.31^{+0.11}_{-0.15}$ and $H_0=69.28 \pm 0.56$ km/s/Mpc ($b = 0.41^{+0.14}_{-018}$ and $H_0=69.70 \pm 0.63$ km/s/Mpc without ACT). 
Finally, we forecast constraints on $b$ for the full SPT-3G survey, Simons Observatory, and CMB-S4, finding improvements by factors of 1.5 (2.7 with Planck), 5.9 and 7.8, respectively, over Planck alone.

\end{abstract}

\maketitle

\section{Introduction}
\label{sec:intro}

Magnetic fields are ubiquitous in galaxies and clusters of galaxies, and there are good reasons to suspect that the universe is magnetized on cosmological scales (see \cite{Durrer:2013pga,Subramanian:2015lua,Vachaspati:2020blt} for reviews). Cosmic magnetism may well be of astrophysical origin, having been generated over the course of the structure formation, but the full story of how this would happen is far from complete. Alternatively, all of the observed fields would be simply explained if a primordial magnetic field (PMF) of a certain strength was already present in the plasma prior to the onset of gravitational collapse. Such fields could have been generated in cosmic phase transitions \cite{Vachaspati:1991nm} or during inflation \cite{Turner:1987bw,Ratra:1991bn} and, if detected, would provide an exciting new window into the early universe. With astrophysical mechanisms being difficult to rule out, only cosmic microwave background (CMB) observations could unambiguously prove the primordial origin of the observed fields.
 
A PMF present in the plasma before and during recombination would leave a variety of imprints in the CMB 
\cite{Crittenden:1993wm,Kosowsky:1996yc,
Jedamzik:1996wp,
Jedamzik:1999bm,
Subramanian:1997gi,
Subramanian:1998fn,
Durrer:1999bk,
Seshadri:2000ky,
Mack:2001gc,
Subramanian:2002nh,
Subramanian:2003sh, Mollerach:2003nq,
Sethi:2004pe,Lewis:2004kg,Chen:2004nf,Lewis:2004ef, Scoccola:2004ke,Kosowsky:2004zh,
Tashiro:2005hc,Kahniashvili:2005xe,Zizzo:2005az,Brown:2005kr,
 Yamazaki:2006bq,Giovannini:2006gz,Kahniashvili:2006hy,
 Giovannini:2007qn,
 Seshadri:2009sy,Caprini:2009vk,
Yamazaki:2010nf,Paoletti:2010rx,Shaw:2010ea,Kunze:2010ys,Cai:2010uw,Trivedi:2010gi,Brown:2010jd,Shiraishi:2010yk,
Shiraishi:2011dh,Trivedi:2011vt,Kunze:2011bp,Paoletti:2012bb,Yadav:2012uz,Pogosian:2012jd,
 Kunze:2013uja,Shiraishi:2013wua,Trivedi:2013wqa,Jedamzik:2013gua,De:2013dra,
 Ganc:2014wia, Ballardini:2014jta,Kunze:2014eka,Kahniashvili:2014dfa,
 Ade:2015cva,Chluba:2015lpa,Ade:2015cva,
 Zucca:2016iur,
Sutton:2017jgr,
Paoletti:2018uic,Pogosian:2018vfr}. In particular, as first pointed out in \cite{Jedamzik:2013gua} and later confirmed by detailed magnetohydrodynamics (MHD) simulations \cite{Jedamzik:2018itu}, the PMF induces baryon inhomogeneities (clumping) on scales below the photon mean free path, enhancing the process of recombination and making it complete at an earlier time. This lowers the sound horizon at last scattering $r_\star$ that sets the locations of the acoustic peaks in the CMB anisotropy spectra, measured with an exquisite precision by Planck \cite{Aghanim:2019ame} and other experiments \cite[e.g.][]{Bennett:2012zja, Dutcher:2021vtw, Choi:2020ccd}. Consequently, this mechanism provides the tightest bounds on the PMF from the CMB, capable of probing fields of  $\sim 0.01- 0.05$ nano-Gauss (nG) post-recombination strength \cite{Jedamzik:2018itu}, well-below the $\sim$nG upper bounds based on other CMB signatures \cite{Ade:2015cva}. As one is entering uncharted terrain in terms of the field strength, there is an actual possibility of detecting the PMF. In fact, the magnetically assisted recombination was recently shown \cite{Jedamzik:2020krr} to be a promising way of alleviating the Hubble tension, discussed further below. This requires a comoving pre-recombination field strength of $\sim 0.05$ nG, which happens to be of the right order to naturally explain all the observed galactic, cluster and extragalactic fields.

The Hubble tension refers to the discrepancy between the value of the Hubble constant $H_0$ determined by fitting the $\Lambda$CDM model to CMB data and $H_0$ determined directly from the slope of the Hubble diagram. The statistical significance of the tension is primarily driven by the difference between the measurement of $H_0=73.2 \pm 1.3$ km/s/Mpc \cite{Riess:2020fzl} using Cepheid calibrated supernova and the Planck best-fit $\Lambda$CDM value of $H_0=67.36 \pm 0.54$ km/s/Mpc \cite{Aghanim:2018eyx}. Other independent measurements tend to re-enforce the tension, with a general trend that all measurements that do not rely on a model of recombination give $H_0$ in the $69$-$73$ km/s/Mpc range \cite{Pesce:2020xfe,Wong:2019kwg,Shajib:2019toy,Harvey:2020lwf,Millon:2019slk,Freedman:2020dne}, while estimates based on the standard treatment of recombination are robustly around $67-68$ km/s/Mpc \cite{Ivanov:2019pdj,Aiola:2020azj,Alam:2020sor}. This might point to a missing ingredient in the model of recombination. As lowering $r_\star$ is what one needs to bring the two groups of measurements closer, the magnetically induced baryon clumping could turn out to be that missing piece of the puzzle.

Adding baryon clumping to the $\Lambda$CDM model creates an approximate degeneracy between the clumping amplitude parameter $b$ and the inferred Hubble constant. Whereas the Planck data by itself prefers zero clumping, the inclusion of the local $H_0$ measurements as a prior results in a 3-4$\sigma$ detection of $b$, while still providing a good fit to Planck and with other cosmological parameters hardly changed from that of the best-fit $\Lambda$CDM model. Reducing $r_\star$, while being a prerequisite for resolving the $H_0$ discrepancy, is not necessarily sufficient by itself, as there is much more information in the CMB temperature and polarization spectra than the locations of the acoustic peaks \cite{Knox:2019rjx}. In particular, the Silk damping and the overall amplitude of polarization are sensitive to modifications of recombination history, along with the balance of power between the small and large scale polarization anisotropies \cite{Zaldarriaga:1995gi}. In this light, it is perhaps remarkable that baryon clumping provides an acceptable fit to the Planck data with $H_0 \sim 70$ km/s/Mpc \cite{Jedamzik:2020krr}. In fact, it is not the worsening of the CMB fit, but preserving the agreement with the baryon acoustic oscillations (BAO) data that prevents achieving an even higher value of $H_0$ in this model, which is a general problem for all solutions of the $H_0$ tension based on lowering $r_\star$ \cite{Jedamzik:2020zmd}. As a byproduct, baryon clumping also slightly relieves the $\sigma_8$ tension, a $\sim$2-3$\sigma$ difference between the values of matter fluctuation amplitude and matter density inferred from weak lensing surveys and CMB constraints within the $\Lambda$CDM model \cite{Heymans2021,DESy3}.

The aim of this paper is to examine the impact of baryon clumping on CMB polarization, focusing on the potentially distinguishing signatures on small angular scales. The small scale temperature and polarization anisotropy spectra were recently measured from the Atacama Cosmology Telescope fourth data release (\actdr) \cite{Choi:2020ccd} and the South Pole Telescope Third Generation (SPT-3G) 2018 data \cite{Dutcher:2021vtw}, and are expected to become more accurate after the release of the Advanced ACTpol \cite{Henderson2016} and SPT-3G full survey \cite{benson2014} data and with future data from the Simons Observatory (SO) \cite{Ade:2018sbj} and CMB-Stage 4 (CMB-S4) \cite{Abazajian:2019eic}. Very recently, \cite{Thiele:2021okz} have performed a similar study, using, however, only the combination of \planck\ and \actdr\ data. They conclude that the addition of \actdr\ strengthens the constraints on clumping compared to \planck\ alone.
Another very recent study \cite{Lee:2021bmn} used the \planck\ data to constrain small-scale isocurvature baryon perturbations, with the conclusion that they cannot alleviate the Hubble tension. Their results, however, do not apply to the baryon clumping induced by PMFs.
While this paper was in its final editing stages, \cite{Rashkovetskyi:2021} also reported constraints on clumping from \planck\ and forecasts for future experiments, using, however, a model with more degrees of freedom compared to ours and not including the \actdr\ and \sptyr\ data. Here, instead, we calculate and examine in detail the constraints from the latest \ACT\ and \SPT\ datasets, alone and in combination with \planck. We test their consistency and robustness against systematic effects, and combine them with other datasets such as the eBOSS DR16 BAO compilation from \cite{Alam:2020sor} and the Pantheon supernovae sample (SN) \cite{Scolnic:2017caz}. To sample the posterior distributions of cosmological datasets we use the \texttt{CosmoMC} code \cite{Lewis:2002ah}, while we calculate the evolution of the recombination history through \texttt{recfast} \cite{recfast}.

The rest of the paper is organized as follows. Sec.~\ref{sec:review} briefly reviews the physics of baryon clumping induced by PMFs and its impact on recombination and CMB polarization signatures that can help distinguish between $\Lambda$CDM and \lcdmb. In Sec.~\ref{sec:separate} we derive constraints on the \lcdmb model from \planck, \actdr\ and \sptyr, and examine the consistency between them investigating the impact of possible systematic effects. In Sec.~\ref{sec:update} we provide joint constraints from CMB, BAO, and SN data. Forecasts of constraints on clumping from SPT-3G, SO and CMB-S4 are given in Sec.~\ref{sec:forecasts}. We conclude with a discussion in Sec.~\ref{sec:discussion}.

\section{Recombination with magnetic fields and the impact on the CMB anisotropies}
\label{sec:review}
We start this section with a review of the physical mechanism behind the magnetically sourced baryon clumping, and the subsequent effect on recombination and the CMB anisotropies. 

\subsection{Recombination with primordial magnetic fields}

A PMF can be generated either during cosmic phase transitions or during inflation. The resultant field is stochastic, but statistically homogeneous and isotropic. The PMF generated in phase transitions would have a very blue spectrum, whereas the simplest inflationary magnetogenesis models predict a scale-invariant PMF. For details on their evolution well before recombination we refer the reader to \cite{Banerjee:2004df}.

PMFs generate baryonic density fluctuations on small scales before recombination. These scales, {\it e.g.} $\sim 1$ kpc for a field of $\sim 0.1$ nG\footnote{All cited length scales and field strengths are comoving.}, are well below the photon mean free path, $l_{\gamma}\sim 1$ Mpc.
The electron-baryon fluid is initially at rest and uniform. The magnetic stress term in the Euler equation, $\propto {\vec{B}}\times({\nabla}\times {\vec{B}})$, induces fluid motions in the plasma, which, via the continuity equation, lead to density fluctuations. The amplitude of the density fluctuations is limited by the back-reaction of the fluid due to pressure gradients. A simple analytic estimate made in~\cite{Jedamzik:2013gua} showed that the amplitude of baryonic density fluctuations follows
\begin{equation}
\frac{\delta \rho_b}{\rho_b} \simeq {\rm min}\biggl[1,\biggl(\frac{c_A}{c_b}\biggr)^2\biggr]
\, ,
\end{equation}
where $c_A = B/\sqrt{4\pi\rho_b}$ is the Alfven speed, with $B$ being the magnetic field strength, and $c_b$ is the baryonic speed of sound,
excluding contributions from photons as they are free-streaming on small scales. Since, at recombination, $c_A = 4.34{\rm km/s}\,(B/0.03nG)$ and $c_b = 6.33 {\rm km/s}$, even fairly weak fields may generate order unity density fluctuations on small scales. This simple estimate has subsequently been confirmed by full numerical simulations~\cite{Jedamzik:2018itu}.

It is important to note that density fluctuations can not exceed unity by much, as the source of the density fluctuations, the PMF, dissipates at some point. Along with the PMF, the produced density fluctuations dissipate as well. But, since all PMFs are described by a continuous spectrum, when the PMF and density fluctuations dissipate on a particular scale, density fluctuations are generated on a somewhat larger scale, where the PMF has not had enough time to dissipate. It is thus unavoidable to have a clumpy baryon fluid shorty before recombination, if a PMF is present. The amplitude of the clumpiness is unknown, as it is determined by the unknown PMF strength. Since the phase transition generated PMFs have a blue spectrum, with most of the power concentrated near the dissipation scale, they shed a significant fraction of their power with each e-fold of expansion, as the peak of the spectrum moves to larger scales. Consequently, they produce larger density fluctuations some time before recombination than immediately before recombination. This is not the case for the scale-invariant PMF.

The magnetically induced inhomogeneities are on scales much too small (i.e. $\ell \sim 10^6-10^7$) to be observed directly in the CMB spectrum. However, the baryon clumping has an indirect effect on the CMB anisotropies. As the recombination rate is proportional to the square of the electron density $n_e$, and since, generally, the spatial average $\langle n_e^2\rangle > \langle n_e\rangle^2$ in a clumpy universe, the recombination rate is enhanced compared to that in $\Lambda$CDM, so that recombination occurs earlier. This, in turn, reduces the sound horizon $r_\star$ at last scattering, the ingredient that many propositions to solve the Hubble tension employ. There is a further, more subtle, effect of the PMF generated baryon clumping on recombination. Local peculiar velocity gradients induced by the PMF influence the local Lyman-$\alpha$ escape rate, with expanding (contracting) regions having a higher (lower) Lyman-$\alpha$ escape rate due to redshifting (blueshifting) \cite{Lewis:2007kz,Senatore:2008vi,Lee:2021bmn}.

The evolution of the cosmic electron density depends on the full shape of the baryon density probability distribution function (PDF), and not only on its second moment, referred to as the clumping factor 
\begin{equation}
b \equiv \biggl(\frac{\langle\rho_b^2\rangle-\langle\rho_b\rangle^2} 
{\langle\rho_b\rangle^2}\biggr).
\end{equation}
The shape of the baryon density PDF is currently unknown, and neither is its evolution before recombination. Furthermore, the statistics of peculiar flows is not known. In the absence of detailed compressible numerical MHD simulations, which would reveal the PDF and peculiar velocity statistics, Refs~\cite{Jedamzik:2013gua} and~\cite{Jedamzik:2020krr} proposed a simple three-zone toy model to estimate the effects of clumping on the CMB anisotropies.
This model computes the average ionization fraction over three different 
regions occupying volume fractions $f_V^i$ and having densities 
$\rho_b^i = \langle\rho_b\rangle \Delta_i$, where $\langle\rho_b\rangle$ is the average baryon density.
Here, somewhat arbritrarily for the model M1, the values $\Delta_1 = 0.1$, 
$f_V^2 = 1/3$, and $\Delta_2 = 1$
are chosen, with the remaining parameters determined by the constraint 
equations
\begin{eqnarray}
\sum_{i=1}^3 f_V^i =1, \ \sum_{i=1}^3 f_V^i\Delta_i = 1, \ 
\sum_{i=1}^3 f_V^i\Delta_i^2 = 1 + b\, .
\label{constraint}
\end{eqnarray}
Modifications to the Lyman-$\alpha$ escape rate are neglected.
We will follow this method here, but {\it alert} the reader to the fact that details of the PDF and its evolution, as well as a modified Lyman-$\alpha$ escape rate, could have an impact on our conclusions regarding
the effects of PMF on the CMB. The current paper rather constrains 
static baryon clumping as a toy model.

\subsection{The impact on CMB anisotropies}
\label{sec:cmbpol}

As discussed previously, baryon clumping facilitates recombination, shifting the peak of the visibility function to an earlier epoch. Along with a smaller $r_\star$, this means that CMB polarization is produced earlier, at a somewhat higher value of the speed of sound $c_S$. As the amplitude of polarization is set by the temperature quadrupole, which is derived from the dipole, which in turn is set by the time derivative of the monopole being proportional to $c_S$ \cite{Zaldarriaga:1995gi}, one generally expects to have a higher polarization amplitude with clumping. More importantly, clumping has a broadening effect on the visibility function, due to overdense baryon pockets recombining earlier and underdense baryon pockets recombining later. The broadening also tends to enhance polarization, because of the longer period of time during which polarization can be generated. 

\begin{figure}[tbph!]
\includegraphics[width=0.48\textwidth]{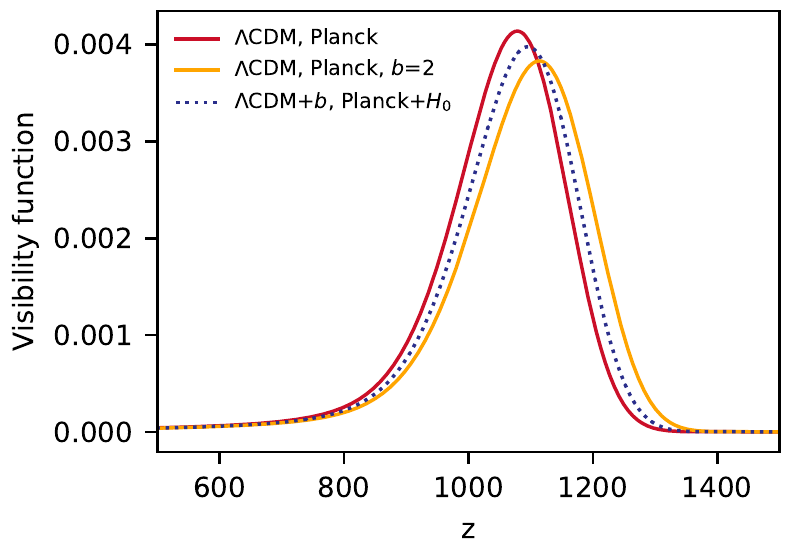}
\caption{Impact of baryon clumping on the CMB visibility function in units of redshift. We show the Planck \LCDM\ best-fit (solid red line), the case where all cosmological parameters are set to the best-fit \lcdm\ and the amplitude of clumping is set to $b=2$ (solid orange line), and the \lcdmb Planck+SH0ES best-fit model (dotted blue line). Clumping shifts the peak of the visibility function to earlier times, and increases its width.}
\label{fig:visibility}
\end{figure}

These effects can be seen in Fig.~\ref{fig:visibility}, which compares the visibility functions\footnote{The visibility function in Fig.~\ref{fig:visibility}, $g(z)$, is the probability density distribution with respect to redshift, {\it i.e.} $dP=g(z)\,dz$. We checked that we obtain a similar impact of baryon clumping on the visibility function when it is defined with respect to conformal time $\eta$ instead of $z$, $g(\eta)=g(z)\,dz/d\eta$.} in the \lcdm\ model, an unrealistic model in which $b=2$ with all other parameters kept the same, and the \lcdmb\ model that best fits the combination of Planck and the SH0ES prior on $H_0$. The broadening effect is apparent from the lower peak, since the plotted visibility function is normalized to integrate to unity. 
This is in contrast to the implicit statement in \cite{Thiele:2021okz} that the visibility function is narrower in clumping models. 

\begin{figure}[tbph!]
\includegraphics[angle=0,width=0.5\textwidth]{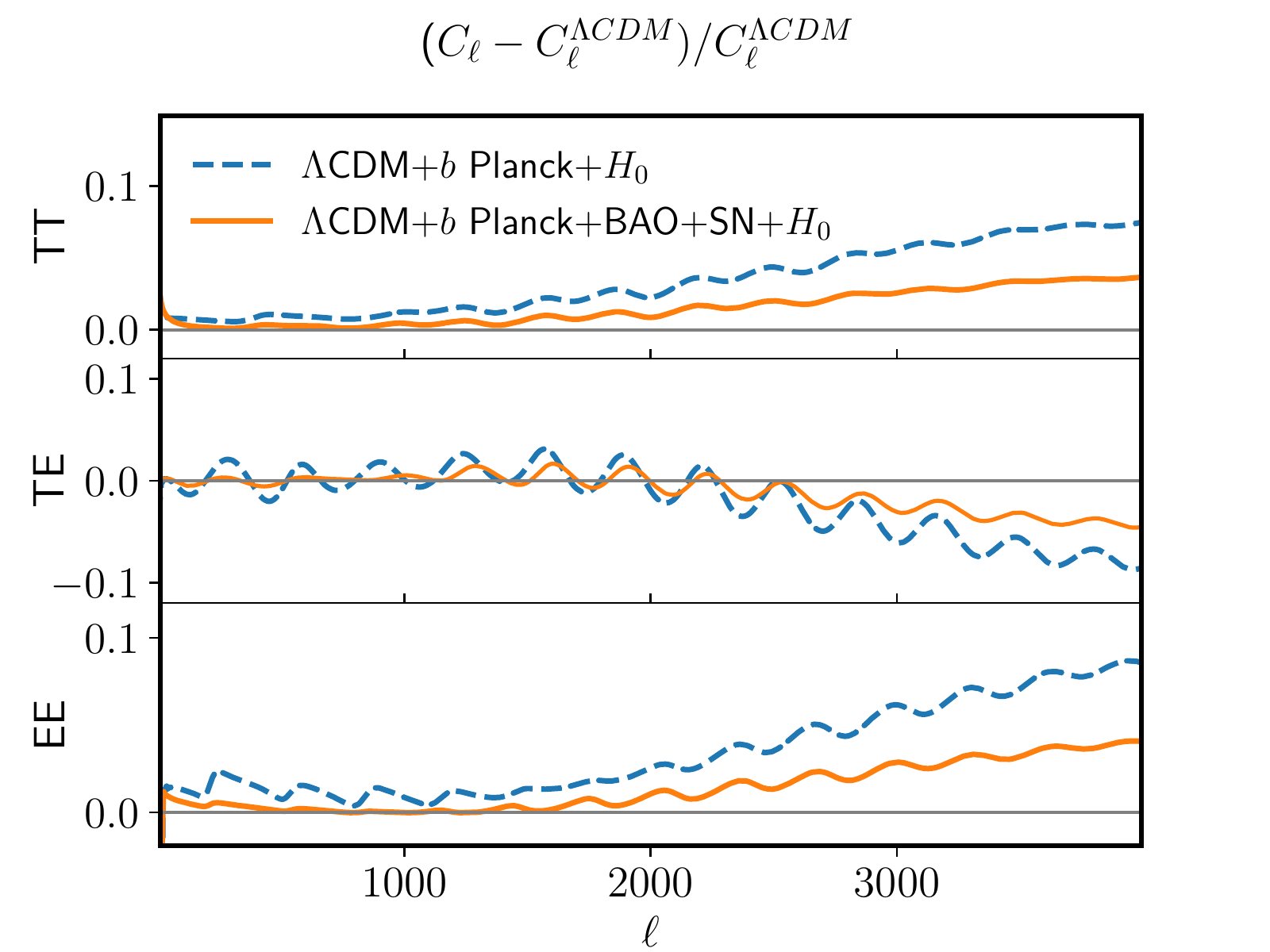}
\caption{Impact of baryon clumping on CMB power spectra. The relative difference between the Planck best-fit $\Lambda$CDM and the M1 model best-fit to Planck+SH0ES (blue dashed line) and Planck+BAO+SN+SH0ES (orange solid line). In the case of TE, to avoid divisions by zero, we compute $(C_\ell - C_\ell^{\Lambda CDM})/C_\ell^{ref}$, where $C_\ell^{ref}$ is the absolute value of $C_\ell^{\Lambda CDM}$ convolved with a Gaussian of width $\sigma_\ell=100$ centred at $\ell$. Baryon clumping leaves signatures in the small-scale anisotropies of the temperature and polarization spectra.}
\label{fig:pmf_cmb_spectra}
\end{figure}

We note that, while the general trends in the visibility function are common to all clumping models, the quantitative details are dependent on the shape and the evolution of the baryon density PDF. The visibility functions shown in Fig.~\ref{fig:visibility} correspond to the particular case of the M1 model, first introduced in \cite{Jedamzik:2013gua}. A second PDF guess, M2, was considered in \cite{Jedamzik:2020krr}, to demonstrate the model-dependence of the results. Because M1 is the more promising of the two for relieving the Hubble tension, all our quantitative results are based on M1.

Another change compared to $\Lambda$CDM is a modification of the Silk damping scale $r_D$. There are three competing effects: $r_D$ decreases due to an earlier recombination, increases due to a smaller electron density before recombination caused by an earlier broad helium recombination, and increases due to the broadening of the visibility function. Note that much of the Silk damping actually occurs right at recombination, where the visibility function is of order unity, such that details of the visibility function matter. The first effect, by itself, would reduce the Silk damping, pushing the onset of the damping tail to higher $\ell$. The second and third effects, however, are also important and are opposite to the first. The balance between them is model-dependent and varies with the clumping factor. In M1, with parameters that fit the data, there is less Silk damping. But in the best-fit M2 model, the Silk damping is virtually identical to that in the best-fit \lcdm. In M1, at (observationally disallowed) high values of $b$, the Silk damping is actually enhanced (see also \cite{Rashkovetskyi:2021} for the the evolution of the damping scale as a function of $b$ in a different PMF model from the one used here).  It should be noted that clumping evolution, which is unknown at present, is yet another source of uncertainty. If clumping was stronger at $z\sim 3000$, then helium recombination could be the dominant effect, inducing more Silk damping. 

Fig.~\ref{fig:pmf_cmb_spectra} compares the CMB spectra in the Planck best-fit $\Lambda$CDM to those in the M1 model that best fits the combinations of Planck+SH0ES and Planck+BAO+SN+SH0ES. We see that both \TT \ and \EE \ are enhanced at high $\ell$ due to reduced Silk damping. At $\ell \lesssim 20$, the polarization is reduced, which is due to the lower best fit value of the optical depth $\tau$. This indicates that high resolution CMB measurements could be a key discriminant in constraining the magnetically sourced recombination.

\section{Constraining clumping with Planck, ACT and SPT-3G}
\label{sec:separate}

As discussed earlier, high resolution CMB temperature and polarization measurements play an important role in constraining the baryon clumping. Hence, it is interesting to investigate the implications of the new data from the ACT and SPT collaborations for this scenario. In what follows, we compare and examine the constraints on clumping from Planck, \actdr\ and \sptyr\ in some detail, including performing tests on simulated data, with the aim of revealing the underlying causes of any differences between them. The datasets we consider are:
\begin{itemize}
	 \item Planck: we use the 2018 final release of the Planck data \cite{Aghanim:2019ame}. At large angular scales, in temperature we use the \TT\ \texttt{Commander} likelihood ($\ell$ = 2-29), while in polarization we use \texttt{SimAll} ($\ell$ = 2-29, \EE\ only). For high multipoles, we use the \texttt{Plik} likelihood in \TT\ ($\ell=30-2508$), \TE\ and \EE\ ($\ell$= 30-1997 in \EE\ and \TE). Finally, we use the \planck\ lensing reconstruction likelihood;
	\item \sptyr: we use the first release of the SPT-3G data described in \cite{Dutcher:2021vtw}. This features \EE\ and \TE\ bandpowers at $\ell= 300-3000$, obtained from observations of 1500 deg$^2$  taken over four months in 2018 (half of a typical observing season) at three frequency bands centered on 95, 150, and 220 GHz. Only about half of the detectors were operable during these observations;
	\item \actdr: we use the fourth release of the ACT data as included in the frequency-combined, CMB-only \actlite\ likelihood \cite{Aiola:2020azj,Choi:2020ccd}, in \TT\,\TE\,\EE. These are based on ACTpol observations taken in 2013-2016 of 6000 deg$^2$ of the sky at 98 and 150 Ghz, as well as the ACT DR2 observations in intensity described in \cite{Das:2013zf};
\end{itemize}
When using the \actdr\ and \sptyr\ likelihoods alone, we use a Gaussian prior on the optical depth to reionization of $\tau=0.0543\pm 0.0073$, following \cite{Aghanim:2018eyx}. When sampling the \lcdmb model, we set a uniform prior on clumping with $0\leq b\leq 2$. Note that since \b\ is weakly constrained in many of the cases considered in the following, a different choice of priors (e.g. a log-uniform one) could have an impact on results. However, we reckon that this choice would not change the qualitative conclusions of this paper.

\subsection{Separate \planck, \ACT\ and \SPT\ constraints}

\begin{figure}[tbph!]
\includegraphics[width=0.48\textwidth]{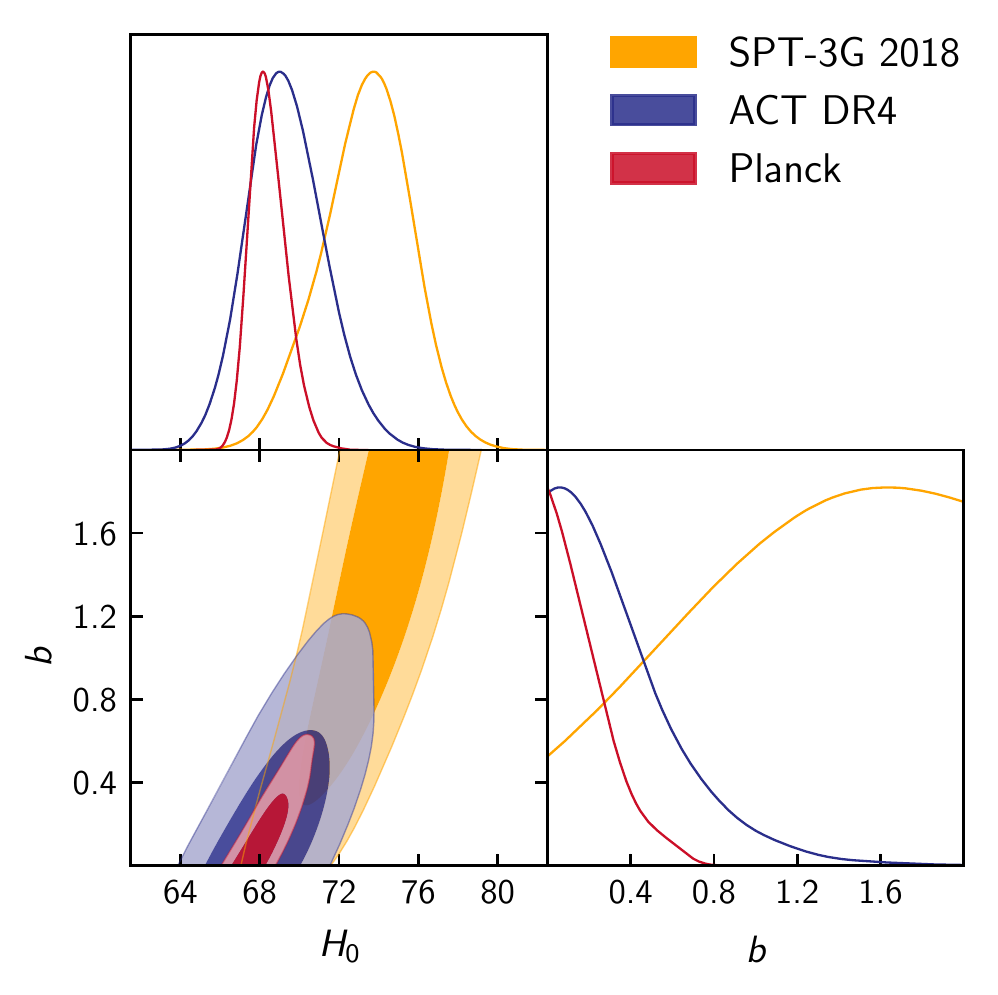}
\caption{Two-dimensional posterior distributions of the clumping factor \b\ and $H_0$ in the \lcdmb\ model for \sptyr\ (yellow), \actdr\ (blue) and \planck\ (red). From \planck\ and \actdr\ we infer best-fit values consistent with no clumping, whereas the posterior for the \sptyr\ data peaks at $b$ larger than one, albeit with low statistical significance.}
\label{fig:act_spt_planck}
\end{figure}

\begin{figure}[tbph!]
\includegraphics[width=0.48\textwidth]{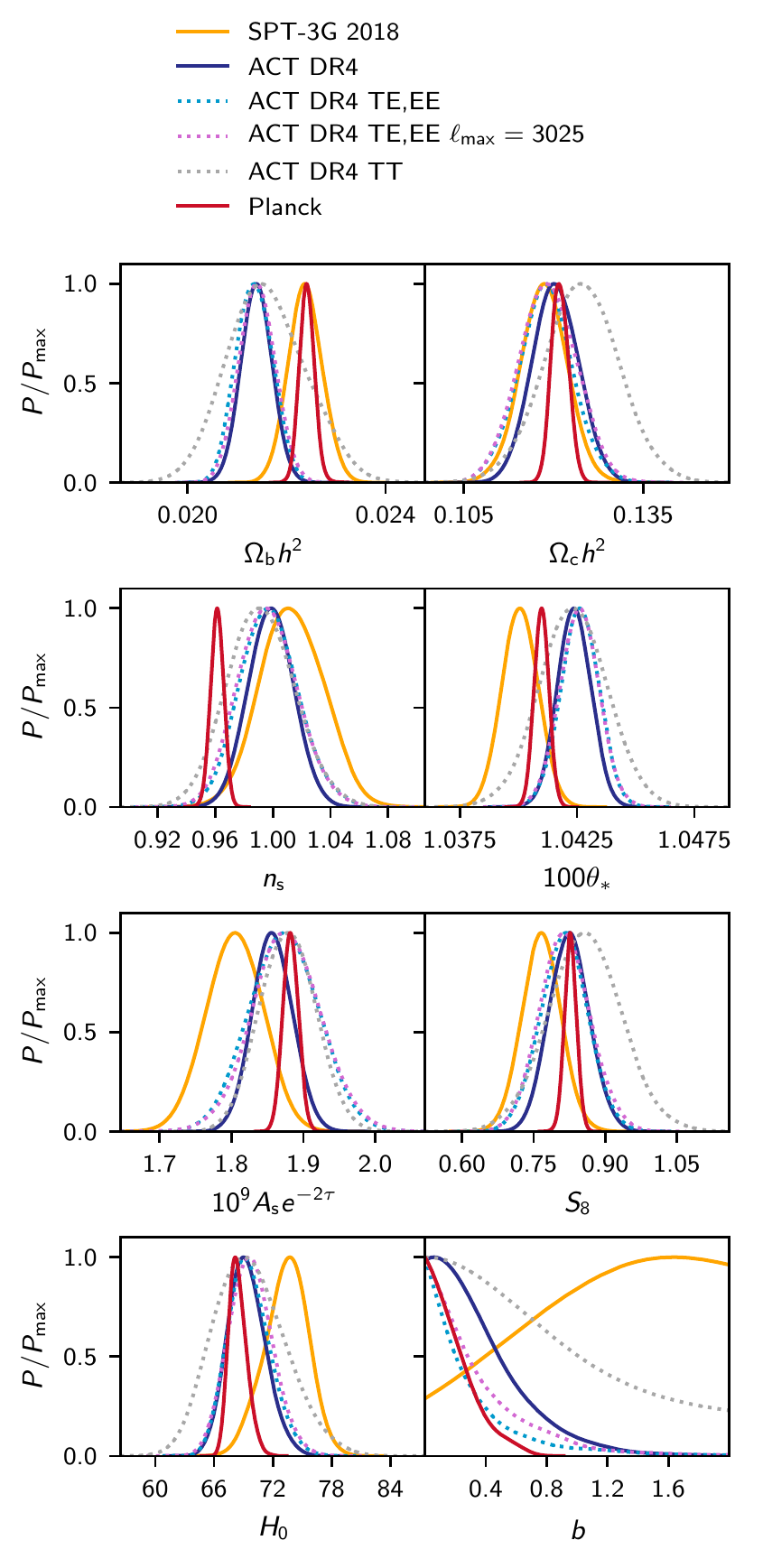}
\caption{One-dimensional posterior distributions of cosmological parameters in the \lcdmb\ model. We show results for \sptyr\ (yellow), \actdr\ (blue),   and \planck\ (red). We also show the impact of separately fitting \actdr\ \TEEE\ (dotted cyan) and \actdr\ \TT\ (dotted gray).} 
\label{fig:act_spt_planck_allpars}
\end{figure}

Fig.~\ref{fig:act_spt_planck} and Fig.~\ref{fig:act_spt_planck_allpars} compare the constraints on the \lcdmb\ model obtained separately from \planck, \actdr\ and \sptyr, also reported in Table~\ref{tab:current}. We highlight three results from these comparisons. First, one can see that the \actdr\ constraint on \b\ is much stronger compared to the one from \sptyr. This is interesting, since for other models, such as \lcdm\ and its extensions considered in \cite{Aiola:2020azj,balkenhol}, the constraints from the two experiments are comparable\footnote{For example, within \LCDM, the ACT constraints are stronger than SPT's only for $\ns$ and $\lnAs$, by $~30\%$. We find this is due to the fact that \ACT\ also includes the \TT\ data, while \SPT\ does not.}.
Second, \actdr\ prefers low values of $\omb$ and high values of $\ns$, as already pointed out for the \LCDM\ model in \cite{Aiola:2020azj}. Third, \sptyr\ prefers values of \b\ different from zero, albeit with a low statistical significance (less than 2$\sigma$). In what follows, we explore in depth the source of these differences.

\begin{table}[tbph]
\def\arraystretch{1.4}
\footnotesize
\setlength{\tabcolsep}{4pt}
\centering
\begin{tabular}{c c c c c c}
 & \planck{} & \sptyr{}  &\actdr{}  &  \makecell{\planck{}\\+\sptyr{}} & \makecell{\planck{}\\+\actdr{}} \\

\hline \hline
$b$ &$<0.51$& $ < 2 $  &  $ < 1.2 $ &$<0.54$ &$<0.31$ \\
$H_0$ &$68.5^{+0.74}_{-1.1}$ & $ 73.4^{+2.4}_{-2.0} $  &  $ 69.3^{+1.7}_{-2.1} $ &$68.7^{+0.76}_{-1.0}$ &$68.1^{+0.63}_{-0.74}$ \\

 \hline
\end{tabular}
\caption{\label{tab:current} Constraints on the clumping \b\ (at 95\% c.l.) and on $H_0$ (at 68\% c.l.) from current CMB data. We note that our adopted prior of $b < 2$ significantly impacts the \sptyr\ constraints on $b$ and $H_0$. Without this prior, \sptyr\ would allow much higher values.}
\end{table}

\subsection{ACT}
\label{sec:act}

\begin{figure}[tbph!]
\includegraphics[width=0.48\textwidth]{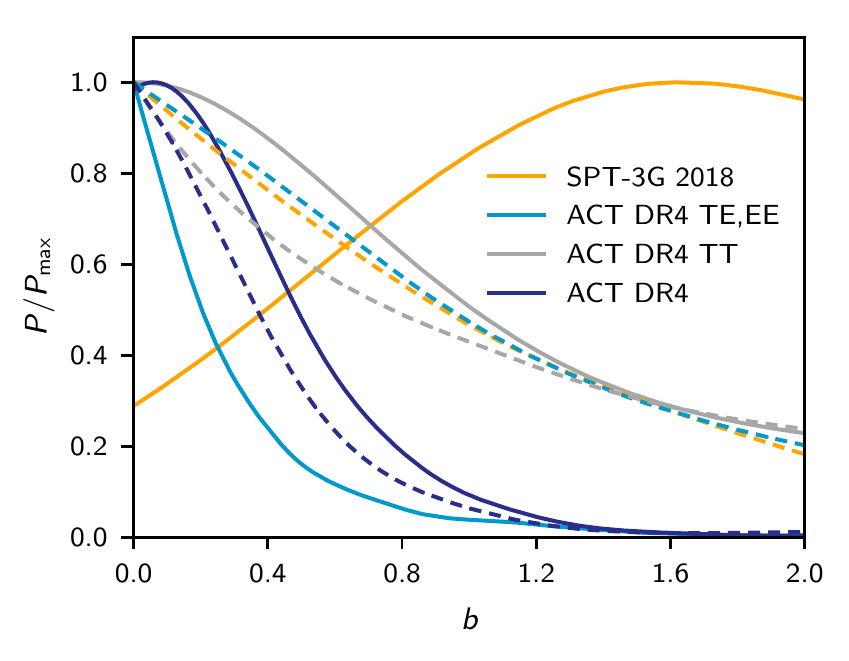}
\caption{Constraints on \b\ from simulated (dashed) and real (solid) data for \sptyr\ (yellow) and \actdr\ (blue).  For \ACT, we also show the results for \TEEE\ (cyan) and \TT\ (gray). 
According to simulations, the tight constraint on \b\ from \ACT\ is due to the combination of \TT\ and \TEEE. Excising the intensity information, \sptyr\ and \actdr\ should have the same constraining power on \b. However, fluctuations in the real data and the physical bound $b>0$ make the \ACT\ \TEEE\ data look much more constraining than that from \SPT.}
\label{fig:act_spt_planck_sim}
\end{figure}

We first investigate the difference in constraining power between \actdr\ and \sptyr. The \actdr\ likelihood includes temperature and polarization \TTTEEE\ spectra  up to multipoles of $\ell_{\rm max}=4000$. On the contrary, the \sptyr\ likelihood includes information from only  \TEEE\ at multipoles up to $\ell_{\rm max}=3000$\footnote{Note that comparing bandpower error bars to investigate the constraining power of the two experiments can lead to wrong conclusions. The bandpowers are correlated -- mostly positively correlated in case of \ACT, while the \SPT\ bandpowers feature strong anti-correlations in adjacent bins. Furthermore, the \actlite\ power spectrum error bars include uncertainties from nuisance parameters, while those of \SPT\ do not, since those parameters are marginalized over at the parameter estimation level.}. 

To assess the expected difference in constraining power, we produce synthetic bandpowers for \actdr\ and \sptyr\, using the \sptyr\ \LCDM\ best-fit as a fiducial model\footnote{The \actdrlite\ likelihood only has only one nuisance parameter, the $\yp$ polarization calibration which multiplies the theory spectra as, {\it i.e.} $C_\ell^{TE}=\yp C_\ell^{th,TE}$, $C_\ell^{EE}=\yp^2 C_\ell^{th,EE}$. In the fiducial model for the synthetic bandpowers, we set $\yp=1$. On the other hand, the \sptyr\ likelihood has several nuisance parameters, which we set to the best-fit values of the \LCDM\ model in the simulations.}. The simulated bandpowers are then analyzed using the same likelihood and nuisance parameters as for the real data. The constraints on \b\ from such simulated datasets are shown in Fig.~\ref{fig:act_spt_planck_sim}.

\begin{figure}[tpbh!]
\includegraphics[width=0.48\textwidth]{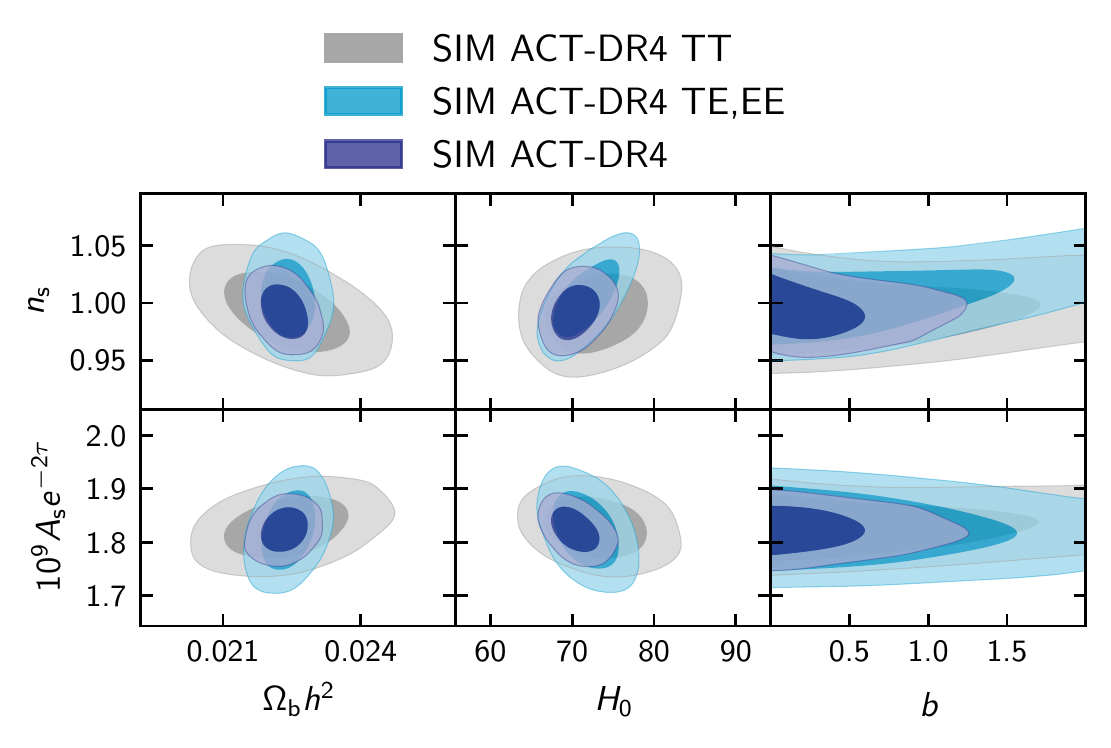}
\caption{Two-dimensional posterior distributions for $\ns$, the amplitude of power spectra $\clamp$ and other cosmological parameters for simulated  \actdr\ bandpowers in the \lcdmb\ model. The simulations assume a $\Lambda$CDM fiducial model with $\b=0$. The combination of \TT\ and \TEEE\ helps to lift degeneracies and to tighten the constraint on \b.}
\label{fig:rectangle_SIM_actns}
\end{figure}

From these simulations, we see that the constraining power of the \actdr\ and \sptyr\ \TEEE\ spectra is practically equal.\footnote{We modified the \actdrlite\ likelihood in order to use the combination of \TEEE\ without \TT. When using \TEEE\ from \ACT, we set a Gaussian prior of $\yp=1\pm 0.01$, in line with the $\mathcal{O}(1\%)$ calibration prior in the \SPT\ likelihood.}
It is the addition of intensity information in the full \ACT\ data that helps to break degeneracies between $\ns$, $\clamp$, and other parameters, which in turn leads to a tighter constraint on \b, as seen in Figure \ref{fig:rectangle_SIM_actns}.

However, the effect of removing \TT\ from the real \ACT\ data is opposite to what is expected from simulations. In particular, the \ACT\ \TEEE\ constraint is {\it stronger} than that from the full \ACT\ likelihood with \TTTEEE, as shown in Fig.~\ref{fig:act_spt_planck_allpars}. We verified that cutting the \ACT\ multipoles at $\ell>\ell_{\rm max}\sim3000$, as done for \sptyr, has a negligible impact on the constraints.

\begin{figure}[tbph!]
\includegraphics[width=0.48\textwidth]{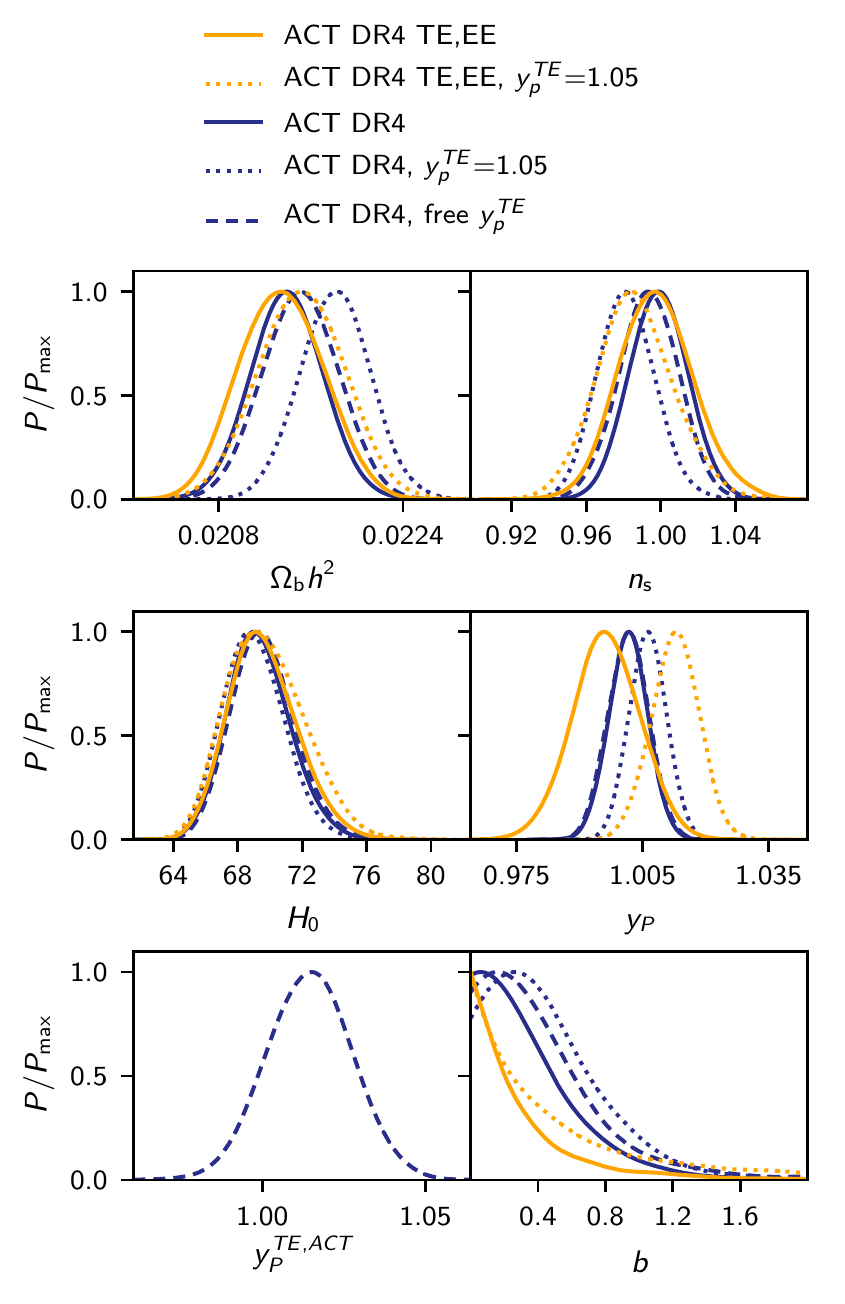}
\caption{Impact of a recalibration of the \ACT\ \TE\ power spectra $\ypTE$ on the \ACT\ \TTTEEE\ (blue) and \ACT\ \TEEE\ (yellow) results. We show the reference case $\ypTE=1$ (solid), the case where we fix $\ypTE=1.05$ (dashed), and the case where we place a uniform prior on $\ypTE$ (dotted). For the \TEEE\ cases, we set a prior on the polarization calibration of $\yp=1.0 \pm 0.01$. For all cases with $\ypTE\neq 1$, we adjust $\yp$ to only affect the \EE\ spectra. A change in the \TE\ calibration weakens the constraints on \b\ and shifts $\ns$ ($\omb$) to lower (higher) values, towards the \planck\ results.}
\label{fig:act_yp}
\end{figure}

As discussed in \cite{Aiola:2020azj}, there are inconsistencies between the \ACT\ and \planck\ results which might be solved by a recalibration of the \ACT\ \TE\ spectra by $\sim 5\%$.  While such a recalibration is not justified by any known source of systematics, we test how this would impact our results. We thus modified the \ACT\ likelihood in order to multiply the \TE\ theory spectra (both for the deep and the wide survey) by a factor $y^{TE}_p=1.05$, while maintaining the baseline $\yp$ polarization calibration parameter for the \EE\ spectra with a Gaussian prior of $\yp=1\pm 0.005$. We find that, indeed, such a recalibration can weaken the constraint on \b\ from \ACT\ \TEEE\ by about 30\%, from $\b<1.0$  (\ACT\ \TEEE) to  $b<1.3$ (\ACT\ \TEEE, $\ypTE=1.05$) at 95\% c.l., as shown in Fig.~\ref{fig:act_yp}. However, the constraint is still stronger than the one expected from simulations by about a factor of $2$, indicating that such a recalibration could only partially account for the difference between the two. On the other hand, the recalibration shifts $\ns$ and $\omb$ constraints towards better agreement with \planck, similar to the \lcdm\ case explored in \cite{Aiola:2020azj}.

Interestingly, the overall \ACT\ \TTTEEE\ constraint is in good agreement with expectations, as shown in Fig.~\ref{fig:act_spt_planck_sim}.
Also in this case, we verify the impact of a recalibration of \TE. We either place a uniform prior on $\ypTE$, or set it to $\ypTE=1.05$. We find that this weakens the 95\% c.l. upper limit on clumping by a smaller amount, from $\b<0.96$ (\ACT\ \TTTEEE) to $\b<1.08$ ($\ypTE$=1.05) or $\b<1.11$ (free $\ypTE$), as shown in Fig.~\ref{fig:act_yp}. Remarkably, the peak of the posterior distribution of $H_0$ is robust against these changes. 

To summarize, the \actdr\ constraints on \b\ are expected to be stronger than those from \sptyr\ because the \ACT\ \TT\ information helps to break degeneracies between $\b$ and other cosmological parameters. Simulations show that \ACT\ \TEEE\ and \SPT\ \TEEE\ should have equivalent power to constrain \b.  However, the constraints based on the real \ACT\ data behave curiously when \TT\ is excised, with the bounds on clumping becoming tighter when only \TEEE\ are used.  While we find that a \TE\ recalibration can impact the \b\ constraints, the \ACT\ \TEEE\ results remain somewhat stronger than what is expected from simulations. At present it is thus unclear whether this is due to a statistical fluctuation or a systematic effect. Overall, the full \ACT\ \TTTEEE\ constraint is in good agreement with expectations.

\subsection{SPT}
\label{sec:spt}

\begin{figure}[h!]
\includegraphics[width=0.48\textwidth]{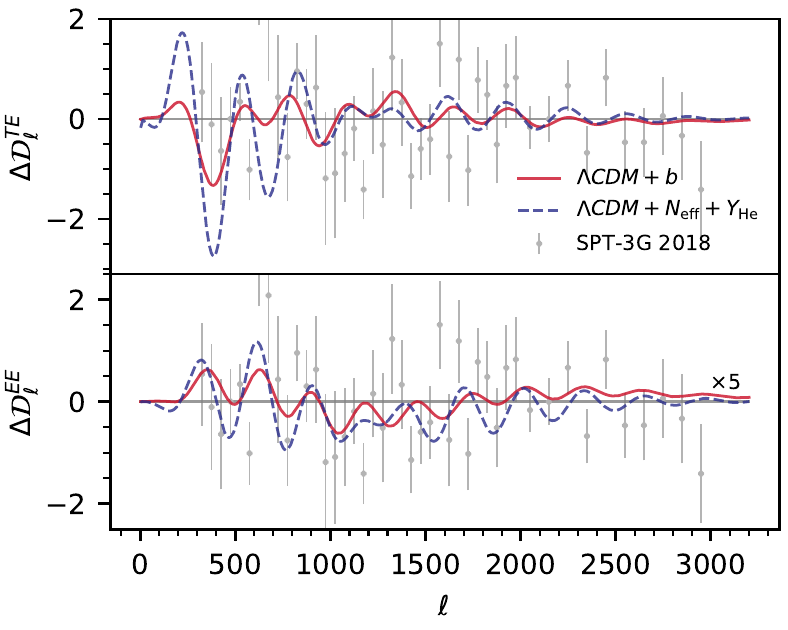}
\caption{Difference in best-fit models of the \SPTyr\ data with respect to \LCDM. We show the case of \lcdmb (solid red) and \lcdm$+\nnu+Y_{\rm HE}$ (dashed blue). The \sptyr\ data residuals with respect to \lcdm\ are shown in gray. The best-fit lines in \EE\ have been multiplied by a factor $5$ in order to allow the comparison with the data points. The \lcdmb and \lcdm$+\nnu+Y_{\rm HE}$ models both fit similar features in the \SPT\ power spectra.}
\label{fig:difference_best_fit}
\end{figure}

Next, we investigate the preference by \sptyr\ for high values of \b. The difference in the best-fit $\chi^2$ with respect to \LCDM\ is only $\Delta_\chi^2=1.7$, thus making this compatible with a statistical fluctuation. We find that this deviation is possibly sourced by the same features in the power spectra that also cause other deviations from \lcdm\ in the \SPT\ results, at a comparable low statistical significance. In particular, \cite{balkenhol} reports a high number of relativist species $\nnu$ and low Helium abundance $Y_{\rm HE}$ compared to the \LCDM\ expectation, when these are fit simultaneously. We find that the \sptyr\ best-fit power spectra for the \lcdmb\ and \LCDM+$\nnu+Y_{\rm HE}$ share key features, as shown in  Fig.~\ref{fig:difference_best_fit}. We cross-checked that for \lcdmb$+\nnu+Y_{\rm HE}$, the deviations with respect to \LCDM\ of all three parameters indeed decrease, confirming the connection between the two models. However, we remind the reader that the improvement in $\Delta \chi^2$ for the two models is not statistically significant: $\Delta \chi^2=1.7$ and $\Delta \chi^2=2$ for one and two additional degrees of freedom for the \lcdmb\ and \LCDM+$\nnu+Y_{\rm HE}$ models, respectively. 

Furthermore, \cite{balkenhol} highlights a slight inconsistency in the mean super-sample lensing convergence $\bar{\kappa}$ measured in the SPT-3G sky patch. The expected \LCDM\ value for this sky region is $10^3 \bar{\kappa}_{\rm \text{\LCDM}}=0\pm 0.45$. The \sptyr\ likelihood marginalizes over $\bar{\kappa}$ with a Gaussian prior set by this \LCDM\ expectation, in order to break the large degeneracy between $\bar{\kappa}$ and the angular scale of sound horizon $\theta$.

However, the combination of  \sptyr\ with \planck, when placing a uniform prior on $\bar{\kappa}$, yields  $10^3 \bar{\kappa}_{\rm SPT-3G}=1.6\pm0.56$. We have tested that using this constraint as a prior on $\bar{\kappa}$ instead of $\bar{\kappa}_{\rm \text{\LCDM}}$ does not impact the \sptyr\ results on \b.

\subsection{Consistency of \planck{}, \ACT{}, and \SPT}

We assess the consistency of the three datasets by comparing their results on cosmological parameters by pairs. In particular, we calculate the parameter $\chi^2_p$ defined as $\chi^2_p=\Delta_p^T \Sigma^{-1} \Delta_p$, where $\Delta_p$ is the difference in the marginalized posterior means obtained by two of the  experiments and $\Sigma$ is the sum of their parameter covariance matrices. We consider the five \LCDM\ parameters $\omb$, $\omc$, $\theta_*$\footnote{$\theta_*$ is the angular size of the sound horizon at last scattering. This is different from the approximated parameter $\theta_\mathrm{MC}$, used in \texttt{cosmomc} to sample the parameter space, and which assumes a \lcdm\ recombination model.}, $\ns$, $\Astau$, together with $\b$. We use the combined amplitude parameter, $\Astau$, since the separate $\As$ and $\tau$ constraints are correlated across the different experiments due to the common use of a \planck{}-based prior on $\tau$ (see \cite{addison2015,planckfeatures} for more details).
This procedure has a number of limitations, offering only an approximated way of assessing consistency. First, it is only valid for independent experiments. While the \sptyr\ and \actdr\ observed sky patches do not overlap and are largely independent, and the \SPT\ and \planck\ correlations can be neglected, there is a correlation between \ACT\ and \planck\ in \TT\ at $\ell<1800$ \cite{Aiola:2020azj}. Furthermore, this procedure assumes that the parameter posterior distributions are Gaussian, so that $\chi^2_p$ can be approximated to have a $\chi^2$ distribution, and this is of course not exactly true for the cases considered. Despite these limitations, we judge that this procedure is good enough to identify major consistency issues between experiments. 

Table \ref{tab:consistency} reports the probability to exceed (PTE) both for the \LCDM\ and the \lcdmb\ models. We judge that datasets with differences larger than 3 Gaussian $\sigma$ (i.e. the number of standard deviations equivalent to the reported PTE for a Gaussian distribution) are not in sufficient agreement to be combined. While we find a good agreement between \SPT\ and \planck, \ACT\ is consistent with either of the two experiments for both models but with large deviations, at the level of $2-3\sigma$ in all cases. This is in agreement with the findings of \cite{Aiola:2020azj}, who evaluated the consistency of \ACT\ and \planck\ for the \lcdm\ model at the $2.7\sigma$ level. We conclude that  \planck, \actdr\ and \sptyr\ are consistent enough to combine them together, although the \ACT\ results are more than $2\sigma$ away from the other two experiments.

\begin{table}[t]
\def\arraystretch{1.4}
\footnotesize
\setlength{\tabcolsep}{5pt}
\centering
\label{tab:consistency}
\begin{tabular}{l D{+}{\%\,}{3.6} D{+}{\%\,}{3.6} }
& \multicolumn{1}{c}{\LCDM{}}   & \multicolumn{1}{c}{\lcdmb{}}   \\ 
\hline
\hline
\planck{}, \sptyr{} & 12+(1.2\sigma) & 6+ (1.8\sigma) \\
\planck{}, \actdr{} & 0.5+  (2.7\sigma)& 1.5+ (2.4\sigma)\\
\sptyr{}, \actdr{} &  0.5+ (2.7\sigma) & 0.8+ (2.6\sigma)\\
\hline
\end{tabular}%
\caption{Consistency between \planck, \ACT\ and \SPT{}. We evaluate the PTE for the \LCDM\ or \lcdmb\ models including $\omb$, $\omc$, $\theta^*$, $\ns$, $\Astau$ (and \b), and report in parenthesis the deviation in units of Gaussian $\sigma$ }
\end{table}

\subsection{Joint \planck, \ACT\ and \SPT\ constraints}

\begin{figure}[tbh!]
\includegraphics[width=0.48\textwidth]{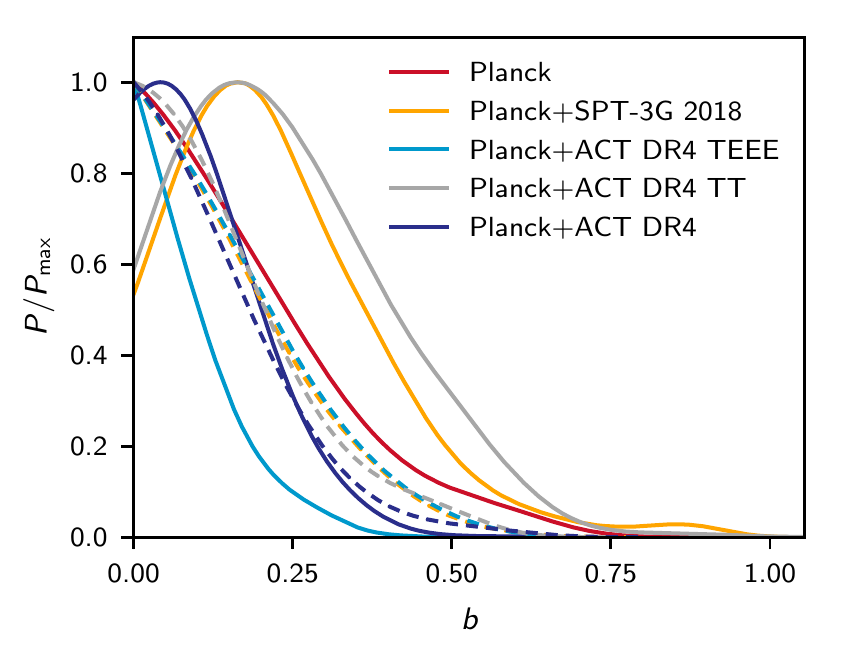}
\caption{Constraints on \b\ from the combination of \planck\ (red) with \actdr\ (blue) or \sptyr\ (yellow). We also show the cases for \planck+\ACT\ \TEEE\ (light blue) and \planck+\ACT\ \TT\ (gray). The solid lines represent results from the real data, while the dashed ones are expectations from simulated bandpowers assuming the \lcdm\ model.}
\label{fig:planckactspt}
\end{figure}

Fig.~\ref{fig:planckactspt} and Table~\ref{tab:current} show the constraints on \b\ when the \planck\ data are combined with \actdr\ or \sptyr. For \ACT, we also show constraints from \TEEE\ and \TT\ separately combined with \planck. We find trends similar to the ones found in Section Sec.~\ref{sec:act}. In particular, the combination of \planck+\ACT\ provides a constraint on clumping, $\b<0.31$ at $95\%$ c.l., which is tighter than the one from \planck+\SPT: $\b<0.54$ at $95\%$ c.l. \footnote{We note that our constraint for \planck+\ACT\ is slightly more stringent that the one reported by \cite{Thiele:2021okz} for the same data combination, $b<0.42$. This might be due to small differences in the implementation of the M1 model, or by the use of different recombination codes, \texttt{HyRec} \cite{hyrec} in \cite{Thiele:2021okz}, \texttt{RECFAST} \cite{recfast} in ours.}

\begin{figure}[tbph!]
\includegraphics[width=0.48\textwidth]{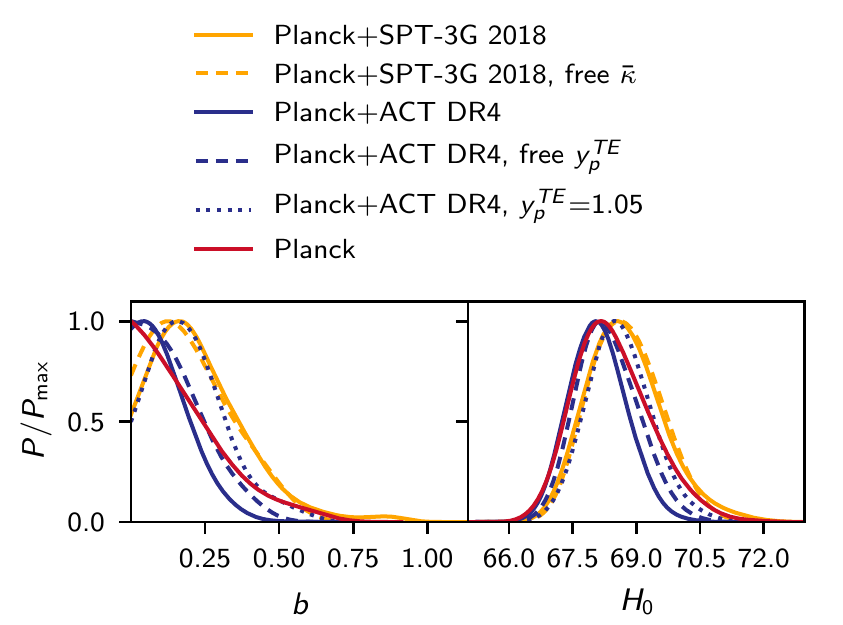}
\caption{Impact of systematics on the constraints on \b\ from the combination of \planck\ with \SPT\ (yellow) or \ACT\ (blue). Placing a uniform prior on the mean lensing convergence for the \SPT\ data (dashed yellow) has a small impact with respect to the reference case (solid). On the contrary, changing the $\ypTE$ calibration of the \ACT\ \TE\ spectra to $\ypTE=1.05$ (dotted) or placing a uniform prior on it (dashed) weakens the constraints.}
\label{fig:planckactsptsystematics}
\end{figure}

Excising the \ACT\ \TT\ information makes the constraint from \planck+\ACT\ \TEEE\ stronger, as already found for the case without \planck. This is in contrast with the results of simulations (already described in Sec.~\ref{sec:act})\footnote{When simulating the combination of \planck\ plus \ACT\ or \SPT, we use the \LCDM\ \planck\ best-fit as a fiducial model for the \ACT\ and \SPT\ simulated bandpowers, while we use the real data for \planck.}, which suggest that  \planck+\ACT\ \TEEE\ and \planck+\SPT\ should provide the same constraint $\b\lesssim 0.40$ at 95\% c.l., as shown in Fig. \ref{fig:planckactspt}. Finally, we find that the combination of \planck+\ACT\ \TT\ fluctuates to values of \b\ higher than zero by $\lesssim 2\sigma$, while from simulations we expect approximately the same constraining power as for \TEEE\ .

We find that the slight preference by \SPT\ for $\b$ larger than zero presented in Sect.~\ref{sec:spt}\ also manifests in the combination with \planck, albeit with very low statistical significance (less than one sigma), providing a constraint of $b<0.54$ at 95\% c.l. Finally, the \planck+\ACT\ results are consistent with expectations and yield $\b<0.31$ ($b<0.39$ 95\% c.l. from simulations).

Similarly to Sec.~\ref{sec:act}, we verify the impact of possible systematics on the constraints, shown in Figure \ref{fig:planckactsptsystematics}. A $5\%$ change in the \ACT\ $\ypTE$ calibration weakens  the \planck+\ACT\ constraint by  $\sim 50\%$, to $\b<0.45$ at 95\% c.l., producing a smaller than $2\sigma$ preference for  $b>0$. Placing a uniform prior on $\ypTE$ has a less dramatic effect, leading to $\b<0.38$ at 95\% c.l. and $\ypTE=1.029\pm  0.014$, i.e. a $2\sigma$ preference for a recalibration in TE. We note that in \LCDM\ we find a similar value, $\ypTE=1.025\pm  0.014$, rather pointing to a $\sim 3\%$ recalibration preferred by the data instead of the 5\% suggested by \cite{Aiola:2020azj}. 

Finally, for \planck+\SPT\ we find that placing a uniform prior on the mean lensing convergence, $\bar{\kappa}$ , across the SPT-3G footprint slightly shifts the peak of the b posterior towards zero, though the upper limit does not change appreciably ($b<0.48$ at 95\% confidence). There is no significant change in the H$_0$ constraint.

\section{The observational status of the clumping model}
\label{sec:update}

Next, we combine the CMB data with the latest BAO, SN and distance ladder measurements, to derive the most up to date constraints on the \lcdmb\ model. Specifically, we use the eBOSS  DR16 BAO compilation from \cite{Alam:2020sor} that includes measurements at multiple redshifts from the samples of Luminous Red Galaxies, Emission Line Galaxies, clustering quasars, and the Lyman-$\alpha$ forest \cite{Zhao:2020tis,Wang:2020tje,Hou:2020rse,duMasdesBourboux:2020pck}, along with the 6dF \cite{Beutler:2011hx} and the SDSS DR7 MGS \cite{Ross:2014qpa} data\footnote{The BAO data-analysis pipelines include the use of \lcdm\ templates which might introduce a bias when analysing the \lcdmb\ model. However, preliminary studies from \cite{Rashkovetskyi:2021} suggest that the impact of \b\ on the BAO templates might be small enough to be ignored.}. We find that the new DR16 BAO data does not make a notable difference compared to the earlier DR12 release when it comes to constraining clumping. For luminosity distances, we use the Pantheon supernovae sample \cite{Scolnic:2017caz}. We also examine constraints with and without the distance ladder estimate of the Hubble constant by the SH0ES team \cite{Riess:2020fzl}, implemented as a gaussian prior of $H_0=73.2\pm 1.3$ km/s/Mpc. We have explicitly checked that using the SH0ES prior on the absolute SN magnitude, as opposed to a gaussian prior on $H_0$, makes no difference for the models we studied. Also, unlike the analysis in \cite{Jedamzik:2020krr}, the SH0ES prior in this work is not combined with the determinations of $H_0$ by the Megamaser Cosmology Project \cite{Pesce:2020xfe} and H0LiCOW \cite{Wong:2019kwg}.

\begin{figure*}[tbph!]
  \centering
  \begin{subfigure}{}
    \includegraphics[width=0.47\textwidth]{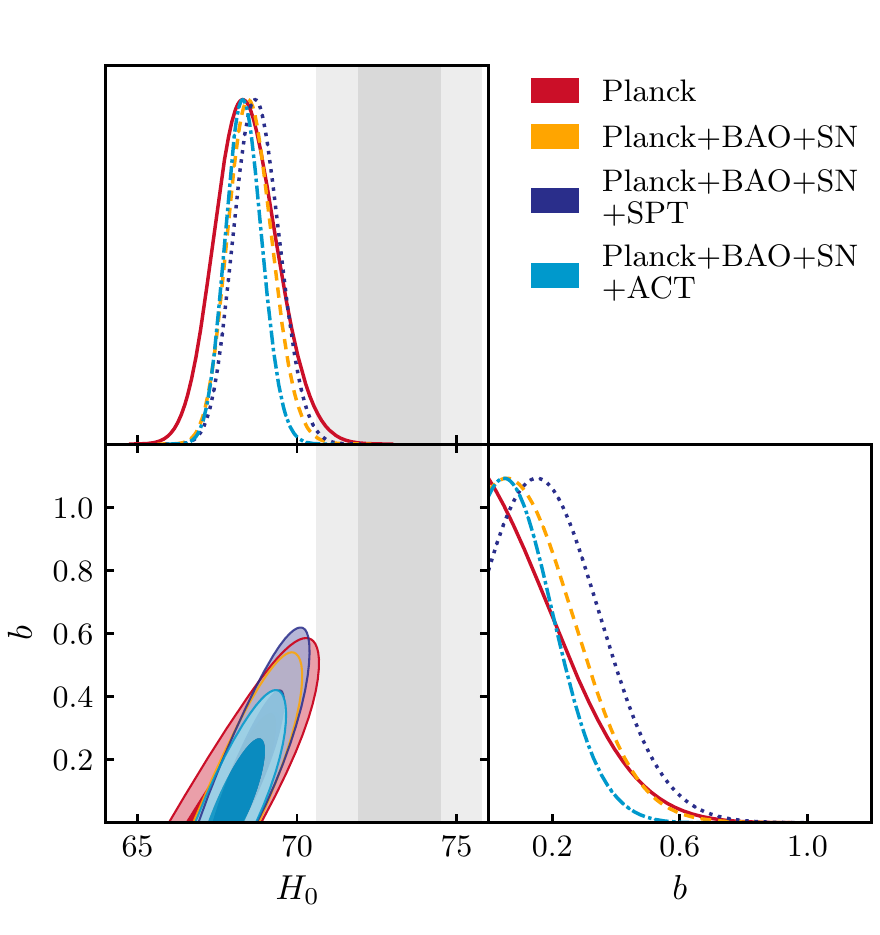}
  \end{subfigure}\hfill
  \begin{subfigure}{}
    \includegraphics[width=0.47\textwidth]{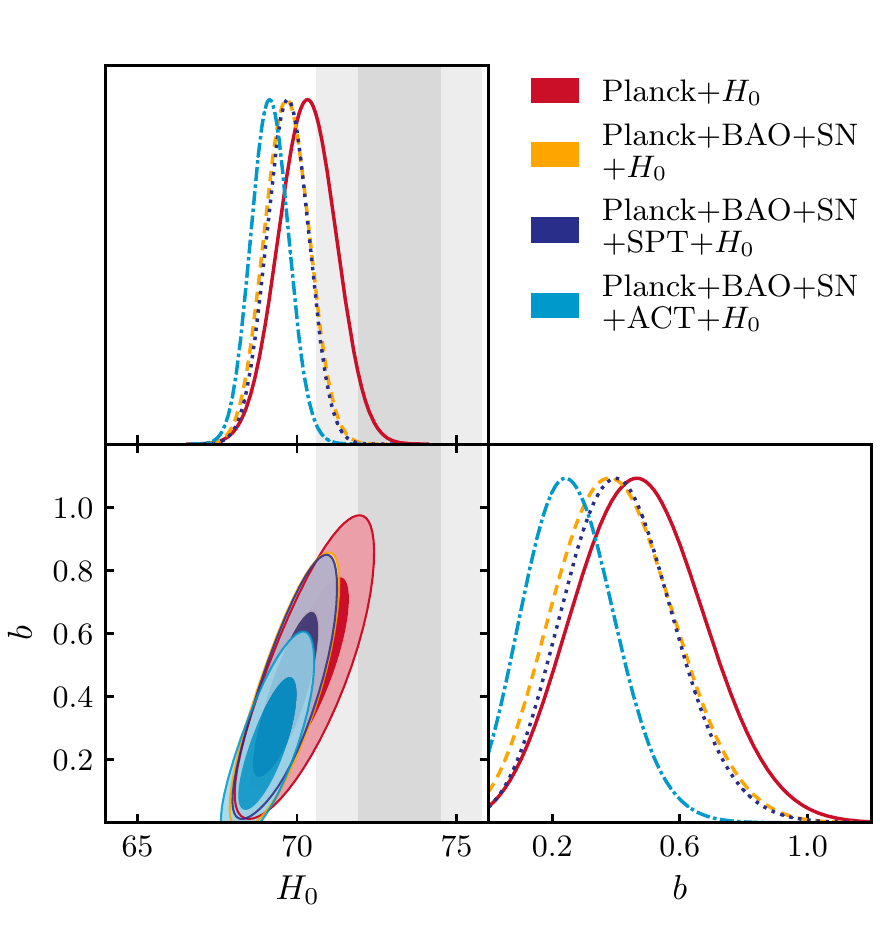}
  \end{subfigure}
  \caption{  \label{fig:update1}
\textit{Left panel:} Constraints on the clumping factor $b$ and $H_0$ from \planck\ only (red), the combination of \planck\ with BAO and SN data (orange), and with \sptyr\ (dark blue) and \actdr\ (light blue). The vertical grey band indicates the 2$\sigma$ range of the latest $H_0$ measurement by the SH0ES team \cite{Riess:2020fzl}. Adding the \sptyr\ data to Planck+BAO+SN relaxes the 95\% c.l. bound on $b$ from $0.43$ to $0.50$, while adding the \actdr\ data tightens it to $b<0.34$. \textit{Right panel:} Same as the left panel, but now including the SH0ES result as a prior on $H_0$.  With the $H_0$ prior, all data combinations show a clear preference for clumping, but  the detection significance is reduced from $\sim$3$\sigma$ to $\sim$2$\sigma$ when the \actdr\ data is included.
}
\end{figure*}

The left panel of Fig.~\ref{fig:update1} shows the marginalized posteriors for the clumping factor $b$ and $H_0$ from Planck, Planck+BAO+SN, Planck+SPT+BAO+SN and Planck+ACT+BAO+SN. We see that Planck by itself shows no preference for clumping, with a small increase in the best-fit $H_0$. Combining Planck with either BAO+SN, \ACT\ or \SPT\ results in a marginal shift of the peaks of the posteriors toward a non-zero $b$. In the case of the BAO+SN, it is due to the mild preference of the BAO data, when analyzed in a recombination-model-independent way \cite{Pogosian:2020ded,Lin:2021sfs}, for a smaller value of the sound horizon at decoupling, $r_{\rm drag}$, and, hence, a larger $H_0$, and the fact that clumping reduces $r_{\rm drag}$.

Adding SPT to the combination of Planck+BAO+SN further increases the peak values of $b$ and $H_0$, while adding ACT has the opposite effect, in line with our earlier results. Adding \ACT\ data results in a notable tightening of the clumping posterior, lowering the upper bound on $b$. The $95$\% c.l. upper bounds on clumping from Planck+SPT+BAO+SN and Planck+ACT+BAO+SN are $b<0.50$ and $b<0.34$, respectively, with additional parameter constraints provided in Table~\ref{tab:mean_params1} in the Appendix. For completeness, the mean parameter values and the 68\% c.l. uncertainties in the \lcdm\ model for the same data combinations are also provided in the Appendix (Table \ref{tab:mean_params0}).

The right panel of Fig.~\ref{fig:update1} shows the impact of adding the SH0ES prior (H0) to the datasets shown in the left panel. Fitting the \lcdmb\ model to Planck+H0 gives $b=0.48 \pm 0.19$ and $H_0 = 70.32 \pm 0.85$ km/s/Mpc. Adding the BAO and SN data results in $b=0.40^{+0.15}_{-0.19}$ and $H_0 = 69.68 \pm 0.67$ km/s/Mpc, a reduction that is generally expected for all models that aim to relieve the Hubble tension by reducing the sound horizon \cite{Jedamzik:2020zmd}. Further combing Planck+BAO+SN+H0 with the SPT data slightly increases the mean values, giving $b=0.41^{+0.14}_{-0.18}$ and $H_0 = 69.70 \pm 0.63$ km/s/Mpc. On the other hand, adding ACT to Planck+BAO+SN+H0 brings the clumping down to $b=0.27^{+0.11}_{-0.15}$, with $H_0=69.15 \pm 0.56$ km/s/Mpc. The global fit to Planck+ACT+STP+BAO+SN+H0 gives $b=0.31^{+0.11}_{-0.15}$ and $H_0=69.28 \pm 0.56$ km/s/Mpc, while reducing the total $\chi^2$ by $5.5$ compared to the \lcdm\ fit to the same data combination. Additional parameter constraints obtained with the SH0ES prior are presented in Table~\ref{tab:mean_params2} in the Appendix. The prior shifts the constraints along the degeneracy axis in the $b$-$H_0$ plane. This results in a clear preference for a non-zero $b$ for all data combinations, though the significance of the clumping detection is reduced from $\sim$3$\sigma$ to $\sim$2$\sigma$ when ACT data is included.

Table~\ref{tab:best_params} (see the Appendix) summarizes the best-fit $\chi^2$ values for the individual datasets in the LCDM and \lcdmb\ models. When comparing the $\Lambda$CDM fits to those of \lcdmb\ for the same data combinations, with and without the SH0ES prior, we see that the $\chi^2$ of Planck (plik), BAO and SN do not change significantly with a non-zero clumping and a higher $H_0$, while the SPT-3G $\chi^2$ is reduced and the ACT $\chi^2$ is increased. This is consistent with the mild tension between SPT-3G and ACT discussed extensively in Sec.~\ref{sec:separate}.

\begin{figure}[tbph!]
\includegraphics[width=0.47\textwidth]{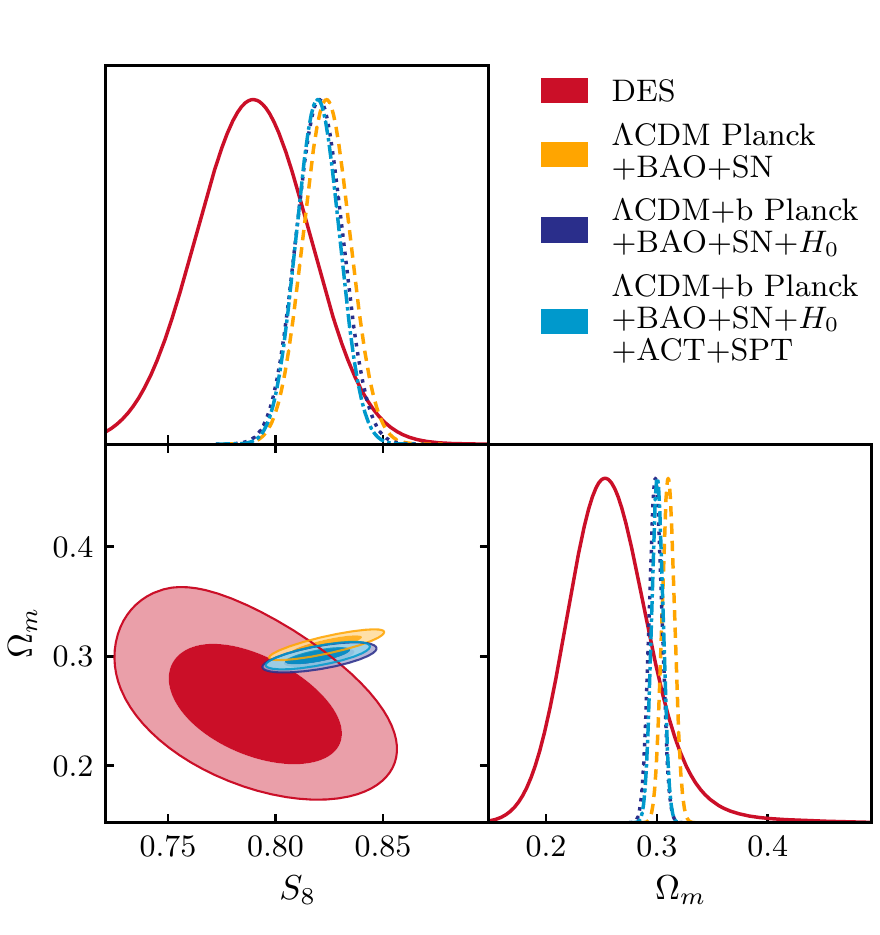}
\caption{\label{fig:s8}Joint constraints in the $S_8$-$\Omega_m$ plane from Planck+BAO+SN in \lcdm\ (orange), in \lcdmb\ with a SH0ES-based $H_0$ prior (dark blue), and when adding \sptyr\ and \actdr\ data to the latter (light blue).
We also show the \lcdm\ based  constraints from the DES Y1 data \cite{Abbott:2017wau} in red. There is a slight relief of the tension in \lcdmb\ due to $\Omega_mh^2$ remaining largely the same as in the best-fit \lcdm{} model, while $h$ is increased.
}
\end{figure}

We note that, in addition to helping to relieve the Hubble tension, the \lcdmb model also improves the agreement between the matter clustering amplitude (quantified by the $S_8$-$\Omega_m$ values) in the Planck best-fit $\Lambda$CDM model and that obtained by the galaxy weak lensing surveys such as DES \cite{Abbott:2017wau,DESy3} and KiDS \cite{Joudaki:2019pmv,Heymans2021}. Fig.~\ref{fig:s8} compares the $S_8$-$\Omega_m$ joint posteriors in the Planck+BAO+SN best-fit \lcdm\ model to those in \lcdmb\ with and without the ACT and SPT data, together with the DES Y1 contours\footnote{The DES Y3 data \cite{DESy3} are in slightly better agreement with \planck\ due to a higher $\Omega_m$. However the DES-Y3 likelihood was not yet available at the time of completion of this paper.}. The primary reason for the lower $S_8$ and $\Omega_m$ values in the clumping mode is the fact that $\Omega_mh^2$ remains largely the same as in \lcdm, while $h$ is increased. 

Summarizing the current status of the \lcdmb\ model, it is clear that it is limited in its ability to fully resolve the Hubble tension, only allowing values of $H_0 \lesssim 70$ km/s/Mpc. However, even if the $H_0$ tension was not fully relieved in this model, a clear detection of clumping is interesting by itself as it would be a tantalizing (albeit indirect) evidence of the PMF. Alternatively, a non-detection of clumping would provide the tightest constraint on the PMF strength.

\section{Forecasts}
\label{sec:forecasts}

In this section, we forecast the constraints on the \lcdmb\ model for ongoing and future experiments. 
We consider three experimental configurations: the full SPT-3G survey \cite{benson2014}, SO \cite{so} and CMB-S4, a next-generation ground-based CMB experiment~\cite{Abazajian:2016yjj}. In all three cases, we consider the combination of the lensed \TTTEEE\ power spectra, while setting a Gaussian prior on the optical depth to reionization of $\sigma(\tau)=0.007$, and do not include the reconstructed CMB lensing spectra. We assume the \LCDM\ \planck\ best-fit, with \b=0, as our fiducial model.

\subsection{Experiments}

\textbf{SPT-3G}. The constraints showed in the previous sections are derived from the SPT-3G \TE,EE\ spectra observed in 2018. This data was collected from only half of a typical observing season, during which half of the detectors were operable. For the forecast presented in this section, we consider five observation seasons (2019--2023 included), which will use all $\sim$16,000 detectors on the main survey field ($\sim$1500 deg$^2$, $f_{sky}\sim 0.03$). We include three frequency bands, 95, 150, and 220 GHz, in both intensity and polarization, with beam full-width half-maximum (FWHM) of 1.7, 1.2, and 1.1 arcminutes (arcmin), respectively, and projected white noise levels in temperature of 3.0, 2.2, and 8.8 $\mu$K-arcmin (a factor of $\sqrt{2}$ higher in polarization)~\cite{benson2014, bender2018}. The noise curves also include the atmospheric $1/f$ noise and account for foreground residuals. The multipole range considered is $\ell=100-5000$.
\newline
\newline
\textbf{SO} is a CMB experiment being built in the Atacama desert in Chile. We use the noise curves for the large-aperture (LAT) 6-m telescope described in \cite{so}, which will observe $40\%$ of the sky ($f_{sky}\sim 0.4$)) in six frequency bands at 27, 39, 93, 145, 225 and 280 GHz. We consider both the baseline and goal sensitivity levels, which correspond to white noise levels in intensity of 8, 10, and 22 $\mu$K-arcmin or 5.8, 6.3, 15 $\mu$K-arcmin at the CMB frequencies of 93, 145 and 225 GHz (a factor of $\sqrt{2}$ higher in polarization), with the beam FWHM of 2.2, 1.4, and 1.0 arcmin, respectively. The noise curves include contributions from the atmospheric $1/f$ noise and foreground uncertainties from component separation (we use the "Deproj-0" configuration from \cite{hill:private}, described in \cite{so}), but also make use of \planck\ data at large angular scales. The multipole range considered is $\ell=40-5000$.
\newline
\newline
\textbf{CMB-S4} is a next-generation ground-based CMB experiment. It will be located in the Atacama desert in Chile and at the South Pole~\cite{Abazajian:2016yjj}, for a wide and a deep area survey, respectively. It will cover frequencies from 20 to 270 GHz. At the main CMB frequencies 93, 145 and 225 GHz, it will feature white noise levels in intensity of 1.5, 1.5, 4.8 $\mu$K-arcmin, with beam FWHMs of 2.2, 1.4, 1.0 arcmin respectively. We forecast the constraints of the wide survey from Chile, using the noise curves from \cite{hills4}, which combine information from all frequencies using an internal linear combination method. Similarly to the SO forecasts, that include the 1/f atmospheric noise and residual uncertainties from component separation. We assume $f_{sky}=0.42$, which excludes the area covering the galaxy in the wide survey.
\newline
\newline
For joint constraints with \planck\, we regard the SPT data as independent.
Due to the small survey footprint of the ground-based survey, we expect correlations to be negligible.
When combining \planck\ with SO or CMB-S4 data, we consider \planck\ data in intensity and polarization at $\ell=30-2500$ covering only the 30\% of the sky which will not be observed by the ground based experiments. Additionally, we use the full Planck data in \TT\ at $\ell<30$, while we do not include polarization at large scales assuming that the information is already contained in the prior we set on $\tau$.

\begin{figure}[tbph!]
\includegraphics[width=0.48\textwidth]{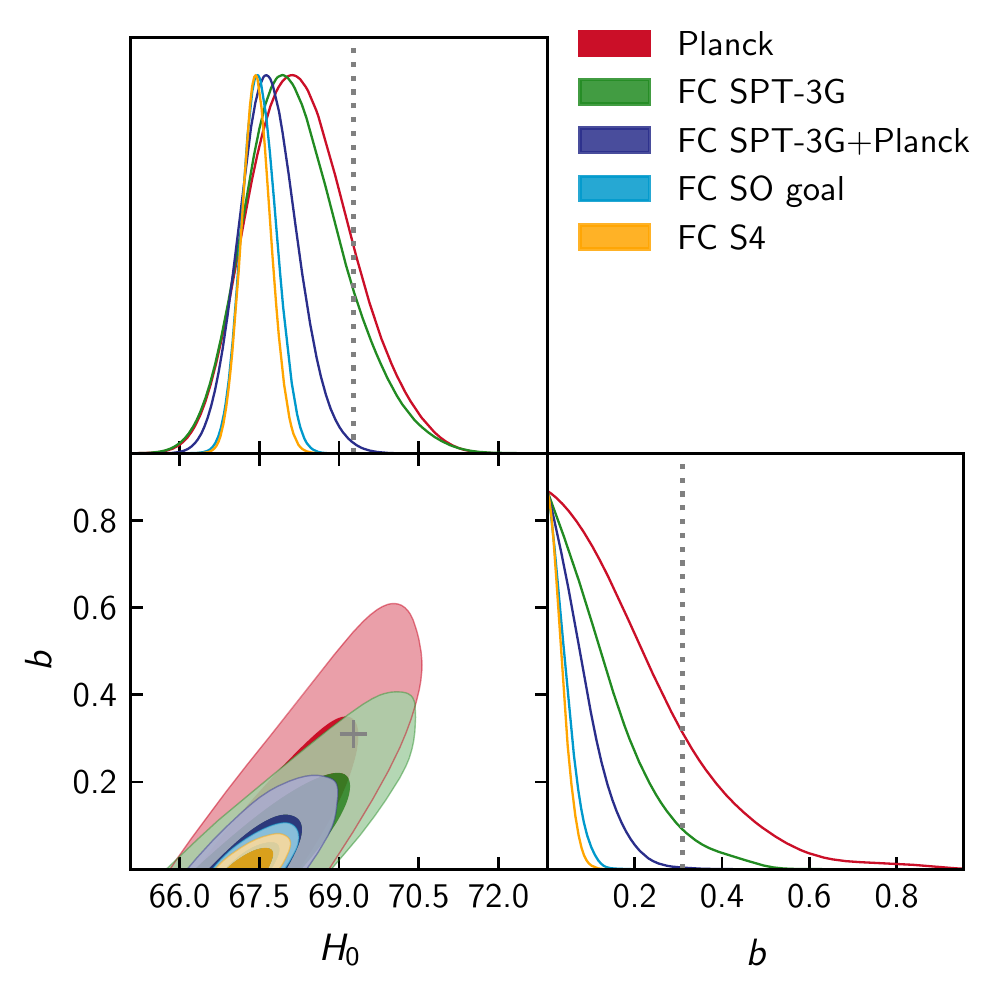}
\caption{Forecasts for \b\ and $H_0$ for the full SPT-3G survey (green) and its combination with \planck\ (blue), for SO (light blue) and CMB-S4 (yellow), assuming a \LCDM\ model as a fiducial. We show the real Planck results (red) as a reference. Future CMB experiments alone will be able to rule out the best-fit of \planck+\ACT+\SPT+BAO+SN+SH0ES in the \lcdmb\ model (gray dotted lines and the cross), {\it i.e.} representing the current ability of the clumping model to relieve the Hubble tension. Conversely, they will be able to detect such a clumping amplitude with some level of confidence.}
\label{fig:forecast1}
\end{figure}

\subsection{Results}

Fig.~\ref{fig:forecast1} shows the results of the forecasts ("FC" in the figures) for SPT-3G, SO (which includes some \planck\ data at large multipoles) and CMB-S4. For SPT-3G, we also show the case where we combine with \planck\ data. We verified that for SO and CMB-S4 adding Planck data does not change the constraints.
We forecast that the full SPT-3G survey will improve the upper limit on \b\ by 50\% compared to \planck, and, in combination with it, by more than a factor of $2.7$. The future generation of CMB experiments will improve the \planck\ limits by a factor of $5.9$ for SO goal (we find no appreciable difference using SO baseline) and  $7.8$ for CMB-S4. We report full results for \lcdmb\ and \lcdm\ in Tables \ref{tab:forecastsb} and \ref{tab:forecasts} in the Appendix, respectively. We note that, while for SPT-3G the constraints on $H_0$ will still weaken by about a factor of 2 in the \lcdmb\ model with respect to \lcdm, the sensitivity of CMB-S4 will break degeneracies sufficiently, so that constraints on $H_0$ and other parameters degrade by $<10\%$ when adding baryon-clumping to \lcdm{}.
In other words, future experiments, such as SO or CMB-S4, will have enough constraining power by themselves
to either detect or rule out the values of clumping currently required to alleviate the Hubble tension. Taking the \planck+\ACT+\SPT+BAO+SN+SH0ES best-fit value of clumping presented in Sec.~\ref{sec:update}, $\b=0.31$ (also shown as a gray cross in Fig.~\ref{fig:forecast1}), \planck+SPT-3G, SO and CMB-S4 will be able to either detect or rule out this value at the $3.5$, $7.5$, $9.7$ $\sigma$ level, respectively. We note that a clumping value of $\b=0.31$ approximately corresponds to a pre-recombination comoving field strength of $B\approx 0.09\,$nG \cite{Jedamzik:2018itu}\footnote{It is important to note that
the limits given in \cite{Jedamzik:2018itu} are on post-recombination magnetic field strength. The relation beween pre- and post-recombination field strength is dependent on the spectral index. Namely, $B_{\rm post}\approx B_{\rm pre}/6$ for phase transition produced fields, while $B_{\rm post}\approx B_{\rm pre}$ for inflationary produced scale-invariant fields. Moreover, due to the use of older 2013 data, the stated limits in \cite{Jedamzik:2018itu} are a factor $\sim 2$ stronger than those from the most recent CMB data.}. 
The most stringent limit on clumping, forecasted for the CMB-S4 experiment, could constrain the pre-recombination PMF field strength to $B < 0.04\,$nG at the $95\%$ c.l.\footnote{The improvement in the constraint on the magnetic field strength $B$ is not
as drastic as that on the clumping factor $b$, as the latter scales as a high power of $B$, {\it i.e.} $b\propto B^x$, with $x\approx 3-4$.}.

\section{Conclusions}
\label{sec:discussion}

Primordial magnetic fields, generated early in the history of the universe, have long been considered as a way of explaining the observed galactic, cluster and intergalactic fields. The evidence for them has grown in recent years, with the blazar observations \cite{Neronov:1900zz,Tavecchio:2010mk,Taylor:2011bn} and the discovery of magnetized filaments of cosmological extent \cite{Govoni_2019}. While there is no firm theoretical prediction for the expected field strength, a pre-recombination PMF of $\sim 0.05$ nG comoving strength would simultaneously explain all observed magnetic fields. It is, therefore, quite intriguing that a PMF of the same strength could also help to relieve the Hubble tension. Indeed, the latter points at a missing ingredient in the physics of the universe at the time of recombination, and many extensions of \lcdm\ were proposed to help resolve the tension (see e.g. \cite{Schoneberg2021}). The baryon clumping induced by the PMF could be that ingredient, with no need to alter \lcdm. 

With no new physics to invent, the PMF sourced clumping is a highly falsifiable proposal. Future CMB experiments, probing temperature and polarization anisotropies at higher resolution, will be able to conclusively confirm or rule it out. There is still uncertainty about the shape and evolution of the baryon PDF.  Obtaining it would require numerical simulations of compressible MHD, which is a challenging, but not impossible, task. Hence, one can fully expect the baryon PDF to be known in due course. Until then, we used a simple model (M1) of clumping, introduced in \cite{Jedamzik:2013gua}, to derive the constraints on the clumping parameter \b\ from the current data, including the recently published CMB data by \actdr\ and \sptyr. We found that the two are in mild tension when it comes to \b, with ACT tightening the Planck bound, and SPT weakly preferring non-zero clumping.

We have investigated potential sources of the difference in constraints. For the \ACT\ data, they appear in part related to the amplitude of the \TE\ spectrum, which has also been shown to cause tension with Planck data within the \lcdm\ model.
Overall, we find that ACT and Planck are in 2.4 $\sigma$ tension in \lcdmb\ compared to 2.7 $\sigma$  in \lcdm, while SPT and ACT are in 2.6 and 2.7 $\sigma$ tension in \lcdm\ and \lcdmb{}, respectively.

Our forecast shows that future high resolution CMB temperature measurements, such as the full-survey SPT-3G data, Simons Observatory, and CMB-S4, will provide a stringent test of this scenario, with the latter capable of constraining \b\ down to 0.065 at 95\% c.l. This corresponds to a pre-recombination PMF strength of 
$\sim$0.04 nG, which would be the tightest constraint on a PMF. 

Baryon clumping, like any other model that aims to reduce the Hubble tension by lowering the sound speed at recombination, can only raise the value of $H_0$ up to $\sim$70 km/s/Mpc, which is still $\sim$2$\sigma$ lower than the SH0ES measurement. The distance ladder measurements of the Hubble constant may still change. However, even if baryon clumping did not fully resolve the Hubble tension, a conclusive detection of evidence for a PMF would be a major discovery in its own right, opening a new observational window into the processes that happened in the very early universe.

\acknowledgments
We are grateful to Karim Benabed, Bradford Benson, Thomas Crawford, Lloyd Knox and the SPT collaboration for insightful discussions and suggestions, and Srinivasan Raghunathan, Marco Raveri, Gong-Bo Zhao for their assistance. We thank Yacine Ali-Ha\"imoud, Julian Mu\~noz, Michael Rashkovetskyi and Leander Thiele for commenting on the manuscript. We gratefully acknowledge using {\tt GetDist} \cite{Lewis:2019xzd}. This project has received funding from the European Research Council (ERC) under the European Union's Horizon 2020 research and innovation programme (grant agreement No 101001897). This research was enabled in part by support provided by WestGrid ({\tt www.westgrid.ca}) and Compute Canada Calcul Canada ({\tt www.computecanada.ca}). This research used resources of the IN2P3 Computer Center (http://cc.in2p3.fr). L.P. is supported in part by the National Sciences and Engineering Research Council (NSERC) of Canada. L.B. acknowledges support from the University of Melbourne.

\appendix

\section*{Appendix}

\label{sec:appendix_data}

We present our full results in the following tables.
We report the constraints on cosmological parameters from the Planck+BAO+SN data with the addition of SPT-3G 2018 and ACT DR4.
In Table \ref{tab:mean_params0} we show results for the \lcdm\ model, whereas we focus on \lcdmb\ in Tables \ref{tab:mean_params1} and \ref{tab:mean_params2} without and with a SH0ES-based prior on $H_0$, respectively.
Table \ref{tab:best_params} shows the best-fit $\chi^2$ values for the models and data combinations considered in Sec.~\ref{sec:update}.
Finally, we present forecast parameter constraints for the \lcdmb\ and \lcdm\ model in tables \ref{tab:forecasts} and \ref{tab:forecasts}, respectively.

\begin{table*}[tbph!]
\def\arraystretch{1.4}
\footnotesize
\setlength{\tabcolsep}{5pt}
\centering
\begin{tabular}{c D{+}{\,\pm\,}{7.7} D{+}{\,\pm\,}{7.7} D{+}{\,\pm\,}{7.7} D{+}{\,\pm\,}{7.7}}          
\multicolumn{1}{c}{}  
& \multicolumn{4}{c}{$\Lambda$CDM Planck+BAO+SN} \\
\cmidrule{2-5}
 & & \multicolumn{1}{c}{+ SPT-3G 2018} & \multicolumn{1}{c}{+ ACT DR4} & \multicolumn{1}{c}{\makecell{+ SPT-3G 2018\\+ ACT DR4}} \\
\hline 
\hline
$\Omega_b h^2$ & 		0.02243+0.00013 & 0.02245+0.00012 & 	0.02240+0.00012 &  0.02242+0.00011 \\
$\Omega_c h^2$  & 		0.11920+0.00090 & 0.11893+0.00086 & 	0.11920+0.00085 &  0.11898+0.00085 \\
$100\theta_{MC}$ & 		1.04101+0.00029 & 1.04079+0.00027 & 	1.04115+0.00027 &  1.04094+0.00025 \\
$\tau$ &  				0.0565+0.0071 & 	0.0563+0.0070 & 	0.0554+0.0070 &  	0.0553+0.0069 \\
${\rm{ln}}(10^{10} A_s)$ & 3.047+0.014 & 	3.045+0.014 & 		3.052+0.014 & 		3.050+0.013 \\
$n_s$ &  				0.9671+0.0036 & 	0.9679+0.0036 & 	0.9693+0.0034 &  	0.9699+0.0033 \\
$H_0$ &  				67.72+0.40 & 		67.76+0.38 & 		67.74+0.38 &  		67.77+0.37 \\
$\Omega_m$ & 		0.3103+0.0054 & 	0.3094+0.0051 & 	0.3100+0.0051 &  	0.3093+0.0050 \\
$\sigma_8$ &  			0.8101+0.0059 & 	0.8084+0.0057 & 	0.8132+0.0057 &  	0.8113+0.0055 \\
$S_8$ &  				0.824+0.010 & 		0.8209+0.0096 & 	0.8267+0.0099 &  	0.8238+0.0095\\
$r_\star$ &  			144.59+0.21 & 		144.65+0.21 & 		144.62+0.21 &  	144.66+0.21 \\
 \hline
\end{tabular}
\caption{\label{tab:mean_params0} Mean parameter values and $68\%$ c.l. uncertainties in $\Lambda$CDM for Planck+BAO+SN and with the addition of the SPT-3G 2018 and ACT DR4 datasets.}
\end{table*}

\begin{table*}[tbph!]
\def\arraystretch{1.4}
\footnotesize
\setlength{\tabcolsep}{5pt}
\centering
\begin{tabular}{c c c c c}          
\multicolumn{1}{c}{}  
& \multicolumn{4}{c}{\lcdmb\ Planck+BAO+SN} \\
\cmidrule{2-5}
 & & + SPT-3G 2018 & + ACT DR4 & \makecell{+ SPT-3G 2018\\+ ACT DR4}  \\
\hline 
\hline
$b$ & 		$< 0.231 (0.434)$ & $0.232^{+0.087}_{-0.19} (0.497)$ & $0.145^{+0.033}_{-0.14} (0.336)$ & $0.175^{+0.063}_{-0.15} (0.378)$  \\
$\Omega_b h^2$ & $0.02242\pm 0.00014$ & $0.02244\pm 0.00012$ & $0.02236\pm 0.00012$ & $0.02237\pm 0.00011$  \\
$\Omega_c h^2$ & $0.1209^{+0.0013}_{-0.0015}$&	$0.1210^{+0.0013}_{-0.0015}$ & $0.1206^{+0.0011}_{-0.0013}$ & $0.1206^{+0.0012}_{-0.0014}$  \\
$100\theta_{MC}$ & 	$1.0452^{+0.0018}_{-0.0034}$ & $1.0459^{+0.0024}_{-0.0033}$ & $1.0445^{+0.0014}_{-0.0027}$ & 	$1.0449^{+0.0018}_{-0.0028}$  \\
$\tau$ & 	$0.0535\pm 0.0072$ & $0.0523\pm 0.0072$ & 	$0.0526\pm 0.0071$ & $0.0518\pm 0.0073$\\
${\rm{ln}}(10^{10} A_s)$ & $3.042\pm 0.014$ & $3.038\pm 0.014$ & $3.046\pm 0.014$ & $3.042\pm 0.014$\\
$n_s$ &  	$0.9618\pm 0.0045$ & $0.9619\pm 0.0044$ & 	$0.9644\pm 0.0043$ & $0.9644\pm 0.0043$\\
$H_0$ & 	$68.54^{+0.56}_{-0.73}$  & $68.74^{+0.61}_{-0.74}$ & $68.34^{+0.48}_{-0.59}$ & $68.48^{+0.52}_{-0.59}$\\
$\Omega_m$ & $0.3065\pm 0.0058$ & $0.3049\pm 0.0056$ & $0.3076\pm 0.0053$ & $0.3063\pm 0.0053$\\
$\sigma_8$ & $0.8162\pm 0.0071$ & $0.8157^{+0.0068}_{-0.0077}$ & $0.8174\pm 0.0063$ & 	$0.8163\pm 0.0064$\\
$S_8$ & 	$0.825\pm 0.010$ & 	$0.8223\pm 0.0098$ & $0.8277\pm 0.0097$ & 	$0.8249\pm 0.0096$\\
$r_\star$ & $143.58^{+0.80}_{-0.51}$ & $143.42^{+0.82}_{-0.62}$ & 	$143.81^{+0.66}_{-0.42}$  & $143.71^{+0.69}_{-0.50}$\\
 \hline
\end{tabular}
\caption{\label{tab:mean_params1} Mean parameter values and along with the 68\% c.l. uncertainties in the \lcdmb\ model for Planck+BAO+SN and with the addition of the SPT-3G 2018 and ACT DR4 datasets. The 95\% c.l. limits on the clumping factor $b$ are given in parenthesis.}
\end{table*}

\begin{table*}[tbph]
\def\arraystretch{1.4}
\footnotesize
\setlength{\tabcolsep}{5pt}
\centering
\begin{tabular}{c c c c c}          
\multicolumn{1}{c}{}  
& \multicolumn{4}{c}{\lcdmb\ Planck+BAO+SN+$H_0$} \\
\cmidrule{2-5}
 & & + SPT-3G 2018 & + ACT DR4 & \makecell{+ SPT-3G 2018\\+ ACT DR4}  \\
\hline 
$b$ &  					$0.40^{+0.15}_{-0.19}$  & 		$0.41^{+0.14}_{-0.18}$ & $0.27^{+0.11}_{-0.15}$ &  		$0.31^{+0.11}_{-0.15}$  \\
$\Omega_b h^2$ &  			$0.02253\pm 0.00014$ & 			$0.02252\pm 0.00012$ & $0.02242\pm 0.00012$ &  		$0.02242\pm 0.00011$ \\
$\Omega_c h^2$ &  			$0.1218\pm 0.0015$ & 			$0.1217\pm 0.0014$ & 	$0.1209\pm 0.0014$ &  		$0.1211\pm 0.0014$ \\
$100\theta_{MC}$ &  		$1.0493\pm 0.0028$ & $1.0493^{+0.0027}_{-0.0025}$ & 		$1.0471\pm 0.0024$ &  		$1.0476\pm 0.0024$ \\
$\tau$ & 					$0.0531\pm 0.0074$ & 			$0.0517\pm 0.0072$  & 	$0.0528\pm 0.0072$ &  		$0.0515\pm 0.0072$ \\
${\rm{ln}}(10^{10} A_s)$ & 	$3.041\pm 0.014$ & 				$3.037\pm 0.014$ & 		$3.045\pm 0.014$ &  		$3.041\pm 0.014$  \\
$n_s$ & 					$0.9603^{+0.0038}_{-0.0043}$ & 	$0.9608\pm 0.0039$ & $0.9632^{+0.0040}_{-0.0045}$ &  	$0.9631^{+0.0038}_{-0.0043}$  \\
$H_0$ & 					$69.68\pm 0.67$ & 				$69.70\pm 0.63$ & 		$69.15\pm 0.56$ &  			$69.28\pm 0.56$ \\
$\Omega_m$ & 			$0.2987\pm 0.0053$  & 			$0.2982\pm 0.0051$  & 	$0.3011\pm 0.0050$ &  		$0.3004\pm 0.0048$ \\
$\sigma_8$ & 				$0.8223\pm 0.0080$ & 			$0.8205\pm 0.0075$ & 	$0.8199\pm 0.0068$ &  		$0.8192\pm 0.0068$ \\
$S_8$ & 					$0.820\pm 0.010$ & 				$0.8179\pm 0.0097$ & 	$0.8214\pm 0.0098$ & 		$0.8197\pm 0.0094$  \\
$r_\star$ & 				$142.71\pm 0.75$  & 			$142.73\pm 0.70$ & 		$143.35\pm 0.63$  &  		$143.20\pm 0.64$ \\
 \hline
\end{tabular}
\caption{\label{tab:mean_params2} Mean parameter values and $68\%$ c.l. uncertainties in \lcdmb\ for Planck+BAO+SN with a SH0ES-based prior on $H_0$ and the SPT-3G 2018 and ACT DR4 datasets.}
\end{table*}

\begin{table*}[tbph!]
\def\arraystretch{1.4}
\footnotesize
\setlength{\tabcolsep}{5pt}
\centering
\begin{tabular}{c D{.}{.}{4.2} D{.}{.}{4.2} D{.}{.}{4.2} D{.}{.}{4.2} c D{.}{.}{4.2} D{.}{.}{4.2} D{.}{.}{4.2} D{.}{.}{4.2} c D{.}{.}{4.2} D{.}{.}{4.2} D{.}{.}{4.2} D{.}{.}{4.2}}
\multicolumn{1}{c}{}
 & \multicolumn{4}{c}{$\Lambda$CDM Planck+BAO+SN} 
 & & \multicolumn{4}{c}{\lcdmb\ Planck+BAO+SN}
 & & \multicolumn{4}{c}{\lcdmb\ Planck+BAO+SN+$H_0$} \\
\cmidrule{2-5} \cmidrule{7-10} \cmidrule{12-15}
 & & \multicolumn{1}{c}{+SPT} & \multicolumn{1}{c}{+ACT} & \multicolumn{1}{c}{\makecell{+SPT\\+ACT}} & \phantom{x} & & \multicolumn{1}{c}{+SPT} & \multicolumn{1}{c}{+ACT} & \multicolumn{1}{c}{\makecell{+SPT\\+ACT}} & \phantom{x} & & \multicolumn{1}{c}{+SPT} & \multicolumn{1}{c}{+ACT} & \multicolumn{1}{c}{\makecell{+SPT\\+ACT}}\\
$\chi^2_{\rm plik}$  & 2346.71 & 2348.71 & 2349.4 & 2350.29 & & 2346.72 & 2350.45 & 2347.6 & 2351.83 & & 2350.98 & 2352.2 & 2348.35 & 	2349.48 \\
$\chi^2_{\rm lowl}$   &  23.60 & 23.37 & 22.28 & 22.32 & & 24.14 & 24.10 & 	23.55 & 22.87 & & 				23.98 & 24.96 & 22.87 & 		23.23 \\
$\chi^2_{\rm simall}$   & 396.71 & 397.27 & 397.95 & 396.05 & & 395.8 & 396.62 & 395.79 & 395.84 & & 		395.68 & 	397.55 & 	395.73  & 	395.78 \\
$\chi^2_{\rm lensing}$   & 8.63 & 8.57 & 8.87 & 	8.85 & & 9.18  & 	8.68 & 8.86 & 8.84 & & 					8.93 & 	8.74 & 	8.86  & 	9.04 \\
$\chi^2_{\rm BAO}$  &  17.58 & 	17.27 & 17.10 & 17.47 & & 17.13 & 17.05 & 	17.31 & 17.20 & &				17.81 & 	18.15 & 	17.03 & 	17.86 \\
$\chi^2_{\rm SN} $  &  1035.08 & 1035.0 & 1034.93& 1035.06 & & 1034.96 & 	1034.92 & 1035.01 & 1034.76 & & 1034.73 & 1034.75 & 	1034.78 & 1034.74 \\
$\chi^2_{\rm SPT}$  &  - & 	1125.76 & - & 1130.01 & & - & 1127.94 & - & 1130.07 & & 						- & 	1124.09 &		 - & 		1126.92 \\
$\chi^2_{\rm ACT} $ &  - & - & 237.56 & 238.67 & & - & - & 241.23 & 238.59 & 	&						- & 		- & 		240.94 & 	242.89 \\
 \hline
\end{tabular}
\caption{\label{tab:best_params} The best-fit $\chi^2$ values for the models and data combinations considered in Sec.~\ref{sec:update}. The SPT-3G 2018 and ACT DR4 datasets are abbreviated as \textit{SPT} and \textit{ACT}, respectively.}
\end{table*}

\begin{table*}[tbph!]
\def\arraystretch{1.4}
\footnotesize
\setlength{\tabcolsep}{5pt}
\centering
\begin{tabular}{c D{.}{.}{2.5} c D{.}{.}{2.5} D{.}{.}{1.2} c D{.}{.}{2.5} D{.}{.}{1.2} c D{.}{.}{2.5} D{.}{.}{1.2} c D{.}{.}{2.5} D{.}{.}{1.2}}
 & \multicolumn{1}{c}{Planck} & \multicolumn{1}{c}{\phantom{x}} & \multicolumn{2}{c}{SPT-3G Y5} & \phantom{x} & \multicolumn{2}{c}{\makecell{SPT-3G Y5\\+Planck}} & \multicolumn{1}{c}{\phantom{x}} & \multicolumn{2}{c}{SO goal} & \multicolumn{1}{c}{\phantom{x}} & \multicolumn{2}{c}{CMB-S4} \\
 \cmidrule{2-2}  \cmidrule{4-5} \cmidrule{7-8} \cmidrule{10-11} \cmidrule{13-14}
 & \multicolumn{1}{c}{$\sigma_p$} & & \multicolumn{1}{c}{$\sigma$}&\multicolumn{1}{c}{$\sigma_p/\sigma$} & & \multicolumn{1}{c}{$\sigma$} &\multicolumn{1}{c}{$\sigma_p/\sigma$} & & \multicolumn{1}{c}{$\sigma$}&\multicolumn{1}{c}{$\sigma_p/\sigma$}& &\multicolumn{1}{c}{$\sigma$} & \multicolumn{1}{c}{$\sigma_p/\sigma$}  \\
\hline \hline
$b$ & <0.5 & & <0.33 & 1.5 & & <0.18 & 2.7 & & <0.085 & 5.9 & & <0.065 & 7.8 \\
$H_0$ & 0.93 & & 0.98 & 0.95 & & 0.56 & 1.7 & & 0.32 & 2.9 & & 0.28 & 3.3  \\

$\Omega_{\mathrm{b}}h^2$ & 0.00015 & & 0.00015 & 1 & & 9.3e-05 & 1.6 & & 5.4e-05 & 2.9 & & 3.8e-05 & 4  \\

$\Omega_{\mathrm{c}}h^2$ & 0.0014 & & 0.0017 & 0.83 & & 0.0012 & 1.2 & & 0.00072 & 2 & & 0.00064 & 2.2  \\

$\tau$ & 0.0073 & & 0.0068 & 1.1 & & 0.0064 & 1.2 & & 0.0055 & 1.3 & & 0.0051 & 1.4 \\

$n_\mathrm{s}$ & 0.0046 & & 0.0079 & 0.58 & & 0.0038 & 1.2 & & 0.0027 & 1.7 & & 0.0025 & 1.9  \\

$\ln(10^{10}A_\mathrm{s})$ & 0.014 & & 0.013 & 1.1 & & 0.012 & 1.2 & & 0.0097 & 1.5 & & 0.0088 & 1.6  \\

$\Omega_{\mathrm{m}}$ & 0.0087 & & 0.012 & 0.75 & & 0.007 & 1.2 & & 0.0043 & 2 & & 0.0038 & 2.3 \\

$S_8$ & 0.013 & & 0.02 & 0.66 & & 0.013 & 1 & & 0.006 & 2.1 & & 0.0048 & 2.7  \\

$\sigma_8$ & 0.0072 & & 0.0068 & 1.1 & & 0.0056 & 1.3 & & 0.003 & 2.4 & & 0.0025 & 2.9  \\

$r_\ast$ & 0.7 & & 0.53 & 1.3 & & 0.37 & 1.9 & & 0.21 & 3.3 & & 0.18 & 3.8  \\
 \hline
\end{tabular}
\caption{\label{tab:forecastsb} Forecasts of constraints on cosmological parameters in the \lcdmb\ model for the full SPT-3G 5-year survey (\textit{SPT-3G Y5}) by itself and jointly with Planck, SO and CMB-S4. 
We also show the error bars from the real \planck\ data for comparison. For each future experiment, the first column shows the  1 $\sigma$ error bars (or the 95\% c.l. upper limit for \b), while the second shows the improvement with respect to \planck\ as the ratio of the uncertainties. Note that the forecasts do not include CMB lensing, which is expected to further contribute to the tightening of the constraints.}
\end{table*}

\begin{table*}[tbph!]
\def\arraystretch{1.4}
\footnotesize
\setlength{\tabcolsep}{5pt}
\centering
\begin{tabular}{c D{.}{.}{2.5} c D{.}{.}{2.5} D{.}{.}{1.2} c D{.}{.}{2.5} D{.}{.}{1.2} c D{.}{.}{2.5} D{.}{.}{1.2} c D{.}{.}{2.5} D{.}{.}{1.2}}
 & \multicolumn{1}{c}{Planck} & \multicolumn{1}{c}{\phantom{x}} & \multicolumn{2}{c}{SPT-3G Y5} & \phantom{x} & \multicolumn{2}{c}{\makecell{SPT-3G Y5\\+Planck}} & \multicolumn{1}{c}{\phantom{x}} & \multicolumn{2}{c}{SO goal} & \multicolumn{1}{c}{\phantom{x}} & \multicolumn{2}{c}{CMB-S4} \\
 \cmidrule{2-2}  \cmidrule{4-5} \cmidrule{7-8} \cmidrule{10-11} \cmidrule{13-14}
 & \multicolumn{1}{c}{$\sigma_p$} & & \multicolumn{1}{c}{$\sigma$}&\multicolumn{1}{c}{$\sigma_p/\sigma$} & & \multicolumn{1}{c}{$\sigma$} &\multicolumn{1}{c}{$\sigma_p/\sigma$} & & \multicolumn{1}{c}{$\sigma$}&\multicolumn{1}{c}{$\sigma_p/\sigma$}& &\multicolumn{1}{c}{$\sigma$} & \multicolumn{1}{c}{$\sigma_p/\sigma$}  \\
\hline \hline
$H_0$& 0.54 & & 0.66 & 0.81 & & 0.47 & 1.2 & & 0.28 & 1.9 & & 0.25 & 2.2 \\

$\Omega_{\mathrm{b}}h^2$& 0.00015 & & 0.00014 & 1 & & 8.9e-05 & 1.6 & & 4.9e-05 & 3 & & 3.7e-05 & 4  \\

$\Omega_{\mathrm{c}}h^2$& 0.0012 & & 0.0017 & 0.69 & & 0.0012 & 1 & & 0.00072 & 1.6 & & 0.00065 & 1.8  \\

$\tau$& 0.0075 & & 0.0068 & 1.1 & & 0.0064 & 1.2 & & 0.0055 & 1.4 & & 0.0052 & 1.5  \\

$n_\mathrm{s}$& 0.0041 & & 0.0076 & 0.54 & & 0.0033 & 1.2 & & 0.0024 & 1.7 & & 0.0022 & 1.8  \\

$\ln(10^{10}A_\mathrm{s})$& 0.015 & & 0.013 & 1.1 & & 0.012 & 1.2 & & 0.0096 & 1.5 & & 0.0089 & 1.6 \\

$\Omega_{\mathrm{m}}$& 0.0073 & & 0.0099 & 0.74 & & 0.0068 & 1.1 & & 0.0042 & 1.7 & & 0.0037 & 2  \\

$S_8$& 0.013 & & 0.018 & 0.7 & & 0.013 & 1 & & 0.0058 & 2.2 & & 0.0044 & 2.9  \\

$\sigma_8$& 0.006 & & 0.0067 & 0.9 & & 0.0055 & 1.1 & & 0.003 & 2 & & 0.0025 & 2.4 \\

$r_\ast$& 0.26 & & 0.44 & 0.6 & & 0.28 & 0.94 & & 0.18 & 1.5 & & 0.17 & 1.6  \\
 \hline
\end{tabular}
\caption{\label{tab:forecasts} Forecasts of constraints on cosmological parameters as in Table~\ref{tab:forecastsb}, but for the \lcdm\ model.}
\end{table*}

\bibliography{polarpmf}

\end{document}